\documentstyle[preprint,aps]{revtex}

\begin{document}

\tighten
\draft

\title{The classification of diagrams in perturbation theory}

\author{D.~R.~Phillips and I.~R.~Afnan}
\address{School of Physical Sciences, The Flinders University of South
Australia,\\ GPO Box 2100, Adelaide 5001, Australia.}

\maketitle

\begin{abstract}
The derivation of scattering equations connecting the
amplitudes obtained from diagrammatic expansions is of interest in
many branches of physics. One method for deriving such equations is
the classification-of-diagrams technique of Taylor. However, as we
shall explain in this paper, there are certain points of Taylor's
method which require clarification. Firstly, it is
not clear whether Taylor's original method is equivalent to the simpler
classification-of-diagrams scheme used by Thomas, Rinat, Afnan and
Blankleider (TRAB). Secondly, when the Taylor method is applied to
certain problems in a time-dependent perturbation theory it leads to
the over-counting of some diagrams. This paper first restates Taylor's
method, in the process uncovering reasons why certain diagrams might
be double-counted in the Taylor method. It then explores how far
Taylor's method is equivalent to the simpler TRAB method. Finally, it
examines precisely why the double-counting occurs in Taylor's method,
and derives corrections which compensate for this double-counting.
\end{abstract}

\newcommand{\be}{\begin {equation}}
\newcommand{\ee}{\end {equation}}
\newcommand{\bea}{\begin {eqnarray}}
\newcommand{\eea}{\end {eqnarray}}
\newcommand{\bel}{\begin {eqnarray*}}
\newcommand{\eel}{\end {eqnarray*}}
\newcommand{\fadj}{f^\dagger}
\newcommand{\bfa}{F^\dagger}
\newcommand{\implies}{\Rightarrow}
\newcommand{\nn}{\nonumber}
\newcommand{\half}{\frac{1}{2}}
\newcommand{\A}[3]{A_{#2 \leftarrow #1}^{(#3)}}
\newcommand{\asm}[3]{a_{#2 \leftarrow #1}^{(#3)}}
\newcommand{\fir}[1]{f^{(#1)}}
\newcommand{\fai}[1]{{f^{(#1)}}^{\dagger}}
\newcommand{\tfai}[1]{{\tilde{f}}^{(#1) \dagger}}
\newcommand{\tpni}[1]{t_{\pi N}^{(#1)}}
\newcommand{\tnni}[1]{T_{N N}^{(#1)}}
\newcommand{\bfi}[1]{F^{(#1)}}
\newcommand{\bfai}[1]{{F^{(#1)}}^{\dagger}}
\newcommand{\cfo}[2]{C_4^{\{#1\}\{#2\}}}
\newcommand{\cfi}[1]{C_5^{\{#1\}}}
\newcommand{\bd}[2]{\bar{\delta}_{#1 #2}}
\newcommand{\ctt}[1]{C_3^{\tilde{t}_i^{(#1)}}}
\newcommand{\cfost}[1]{C_4^{\tilde{s}_f^{(#1)} \tilde{t}_i^{(#1)}}}
\newcommand{\cfs}[1]{C_5^{\tilde{s}_f^{(#1)}}}
\newcommand{\tit}[1]{\tilde{t}_i^{(#1)}}
\newcommand{\sft}[1]{\tilde{s}_f^{(#1)}}
\newcommand{\comb}[2]{\left(  \begin{array}{c}
																															#1\\
																											    #2
																														\end{array}   \right)}
\newcommand{\qeq}{\stackrel{?}{=}}

\section {Introduction}

\label {sec-intro}

The method of classification of diagrams, developed by Taylor, is a
powerful technique by which  equations connecting the amplitudes
obtained from a field theory may be derived, without it being
necessary to explicitly specify the Lagrangian of the theory in the
derivation~\cite{Ta63,Br69}. This model-independence has made the
technique particularly useful in theories of mesons and baryons, where
it is not practical to use the QCD Lagrangian and the best equivalent
Lagrangian containing meson and baryon degrees of freedom is not yet
known. Examples of the application of the technique to simple systems
of nucleons and pions include the original work on equations for the
$\pi \pi\pi$, $\pi \pi N$, $\pi N N$ and $N N N$ systems by Taylor himself
\cite{Ta66}; the derivation of the $\pi N-\pi \pi N$ equations by Afnan and
Pearce~\cite{AP86,AP87}; studies of pion photoproduction on both a
single nucleon and the deuteron~\cite{AA87,Af88,AA88} and the derivation by
Avishai and Mizutani~\cite{Mi76,AM83}, on the one hand, and Thomas,
Rinat, Afnan and Blankleider on the other~\cite {Th73,TR79,AB80}, of
the
$NN-\pi NN$ equations. This work on the $NN-\pi NN$ system raised at
least two questions about the Taylor technique, both of which, despite
the technique's widespread application, remain unanswered. Although
these problems originally arose in the context of the
$NN-\pi NN$ equations it should be clear that the questions themselves
are quite general ones about the Taylor method and, as such, are
relevant independent of the particular system and Lagrangian under
consideration.

The first question arose because Taylor's original technique was
somewhat modified and simplified by, first, Thomas and Rinat \cite{TR79} and,
second, Afnan and Blankleider~\cite{AB80,AB85}, in order to make it
more useful for time-ordered perturbation theory calculations.
However, the equations obtained by Afnan and Blankleider for the
$NN-\pi NN$ system~\cite{AB80} were exactly the same as those derived
by Avishai and Mizutani\footnote{Note that Avishai and Mizutani included both
a term to account for heavy-meson exchange and a three-body force in their
calculation, whereas Afnan and Blankleider included neither of these effects.
However, the heavy-meson exchanges and a three-body force can easily be
included in Afnan and Blankleider's derivation, and when that is done the
resulting equations are exactly those obtained by Avishai and Mizutani}, who
used Taylor's original technique and a
{\em time-dependent} perturbation theory~\cite{AM83}. Is this pure coincidence,
or did Thomas and Rinat and Afnan and Blankleider (TRAB) discover a
simplification of Taylor's technique? This question was posed and left
unanswered by Avishai and Mizutani~\cite{AM83}. In this paper we
answer it by explaining what assumptions must be made if the TRAB
method is to produce the same equations as Taylor's original
technique.

The second problem is that Taylor's method can lead to the
double-counting of certain diagrams when it is applied in a
time-dependent perturbation theory, such as covariant perturbation
theory. This problem was first pointed out by Kowalski, Siciliano and
Thaler who showed that there was double-counting in some models of
pion absorption on nuclei~\cite{Ko79}. While Kowalski et al.\ did not
refer specifically to Taylor's method, the double-counting problem
certainly arises when one applies the classification-of-diagrams
technique to the problem of summing all possible diagrams contributing
to, say, pion absorption on the deuteron. If one applies the Taylor
method, as described below, to pion absorption on the deuteron, one
obtains contributions from both of the diagrams in
Fig.~\ref{Fig.1}. In this figure $t^{(1)}$ is the $\pi N$ t-matrix with the
$s$-channel pole part removed, $T$ is the full $N N$ t-matrix (provided one
assumes the absence of anti-nucleons in the deuteron) and the nature of the
$\pi NN$ vertex is explained below. Kowalski et al.\ pointed out that the
inclusion of the crossed term (depicted in Fig.~\ref{Fig.2}) in $t^{(1)}$ leads
to double-counting, as follows. In a time-dependent perturbation theory the
contribution of this part of $t^{(1)}$ to the diagram on the right of Fig.
\ref{Fig.1} is the diagram shown in Fig.~\ref{Fig.3}. However, this diagram has
already been included as distortion in the initial channel, via the diagram on
the left of Fig.~\ref{Fig.1}.  Note that in a time-ordered perturbation
theory this double-counting problem does {\em not} arise, since the
only contribution made by Fig.~\ref{Fig.2} to the right-hand diagram of
Fig.~\ref{Fig.1} is the diagram shown in Fig.~\ref{Fig.4}. In a time-ordered
approach this diagram is not included as distortion in the incoming $NN$
channel and so is not over-counted.

The existence of this problem raises two questions. Firstly, why does this
erroneous double-counting occur in a method which Taylor claimed worked {\em
regardless of the perturbation scheme} being used? Secondly, how can the
over-counting be eliminated? Avishai and Mizutani attempted to provide answers
to both of these questions in their 1983 paper~\cite{AM83}. They claimed that
the double-counting problem occurred because the derivation of the $NN-\pi NN$
equations had only considered the $s$-channel structure of the  amplitudes in
question. (We are using the notation of Mandelstam here\cite{Ma58,Ma59}.) They
suggested that examining the $s$-, $t$- and $u$-channel structure
simultaneously
would remove the double-counting. However, this proposal contradicts Taylor's
original work in which he derived the classification-of-diagrams technique so
as to have no double-counting {\em whichever channel or channels the
amplitudes'
structure was examined in}. He only proposed performing the structure
examination in a  number of channels simultaneously as part of an
approximation he intended to use in order to close the set of equations
obtained
from his method. We shall see below that Avishai and Mizutani's suggestion is
partly right: the lack of specification of the $t$- and $u$-channel
cut-structure of the sub-amplitudes from which the full amplitude is
constructed can be thought of as the cause of the double-counting problem.
Consequently, if the double-counting is to be eliminated the cut-structure of
these amplitudes in channels other than the $s$-channel does need to be
considered. But, to cut in all channels simultaneously is unnecessary.
Furthermore, such a solution is impractical, as it results in highly non-linear
equations.

Another ``solution" suggested by Avishai and Mizutani is just to ignore the
double-counting, since (they claim) ``\ldots compared with the important role
played  by the direct nucleon pole term in nuclear $\pi$ absorption, the
possible over counting of the crossed pole term should hardly affect the
essential physics."~\cite{AM83}. The accuracy of this statement is open to
question and the validity of such an approach was never tested, since the
numerical calculation based on Avishai and Mizutani's original work used
Blankenbecler-Sugar reduction~\cite{BbS66} in order  to reduce the
four-dimensional equations, thus time-ordering them and eliminating the
double-counting difficulty~\cite{La87}. Therefore this ``solution" is
practical, but, since it amounts to ignoring the problem completely we question
whether it really is a solution at all!

The example presented by Kowalski et al.~\cite{Ko79} shows that
double-counting can occur when the Taylor method is applied to a
time-dependent perturbation theory of the $NN-\pi NN$ system. The presence of
this double-counting indicates a fundamental flaw in the Taylor method, which
must be resolved  before the method can be used confidently in order to derive
four-dimensional equations for {\em any} system. In this paper we solve this
double-counting problem in general, by first pointing out why the
double-counting arises in Taylor's method, and then demonstrating how
correction terms can be introduced to eliminate it.
Consequently, we answer both the questions which were posed about the
double-counting problem above.

Therefore, this paper resolves two issues associated with Taylor's
method: the validity of TRAB's simplification  of Taylor's original
work and the double-counting problem. In order to do this we first
recapitulate Taylor's original argument, in Section
\ref{sec-Taylorrev}. In so doing we find that certain instances of
double-counting arise if Taylor's method is not applied
carefully. However, we show that this type of double-counting may be
eliminated if we constrain the cut-structure of amplitudes in
channels other than the $s$-channel. Then, in Section
\ref{sec-ABTaylor} we compare Taylor's result for an arbitrary
amplitude with the result obtained from TRAB's conceptually simpler
approach. In Section \ref{sec-Taylorwrong} two examples of
double-counting, including the example originally provided by Kowalski
et al.~\cite{Ko79}, are given. The flaw in Taylor's method which leads
to these two instances of double-counting is then discussed, as is the
way in which this type of double-counting differs from that discussed
in Section~\ref{sec-Taylorrev}. Finally, in Section \ref{sec-Dcsoln} a
general solution to this second type of double-counting problem is
provided.

\section {The classification-of-diagrams method of Taylor: a review}

\label {sec-Taylorrev}

In order to place the two points made in this paper about the
classification-of-diagrams technique in their proper context we first
review the Taylor method, summarizing the arguments presented in
Taylor's original paper~\cite{Ta63}. In this review we examine the
method as applied in the
$s$-channel, but with appropriate modifications the technique may be
applied in any channel.

Taylor's method is a topological procedure which allows the summation
of a series of diagrams via the classification of these diagrams
according to their irreducibility. The method does not assume that
these diagrams have been generated by a field theory. The diagrams
could, for example, be diagrams representing the perturbation series
expansion for an interacting system of $N$ particles. However, in this
paper we take the view that the diagrams under consideration are
Feynman diagrams generated by some field theory. In this view, the
Taylor method provides a means of deriving equations connecting the
amplitudes obtained from this underlying field theory.

Therefore, we assume there exists some perturbation expansion of
Feynman diagrams, which when summed give a set of $m
\rightarrow n$ Green's functions, which in momentum-space we represent
by:
\begin{equation}
G_{n \leftarrow m}(p_1',p_2',\ldots,p_{n}';p_1,\ldots,p_m).
\end{equation}
If, of the $m (n)$ particles in the initial (final)
state, $j$ ($j'$) are nucleons and the rest are pions, LSZ reduction
\cite{LSZ55} may be used to obtain the amplitude corresponding to this
Green's function:
\begin{eqnarray}
A^{(j',j)}_{n \leftarrow m} (p_1',\ldots,p_{n}';p_1,\ldots,p_m)d_N^{-1}(p_1')
\ldots d_N^{-1}(p_{j'}') d_\pi^{-1}(p_{j'+1}') \ldots
d_\pi^{-1}(p_{n}')\nn\\
\times\ G^{(j',j)}_{n \leftarrow m}(p_1',\ldots,p_n';p_1,\ldots,p_m)\
 d_N^{-1}(p_1) \ldots d_N^{-1}(p_{j}) d_\pi^{-1}(p_{j+1})
\ldots d_\pi^{-1}(p_{m})\ .                           \label {eq:LSZ}
\end{eqnarray}
(Note that the use of the terms ``initial" and ``final" state here, and in the
ensuing argument, is slightly liberal, since in time-dependent perturbation
theory there is nothing which restricts the times associated with the
$m$-particles  with reference to those associated with the $n$-particles. But,
by ``initial" state we mean the state with $m$ particles having momenta
$p_1,\ldots,p_m$ and by ``final" state we mean the state with $n$ particles,
having momenta $p'_1,\ldots,p'_{n}$.) Taylor's method provides a way of
classifying all the perturbation diagrams contributing to $A_{n \leftarrow m}$
according to their topology.

However, Taylor's method works only if all the particles involved are
fully dressed. In order, therefore, for us to be able to discuss the
Taylor method, we need to assume that all particles are fully dressed.
(For a discussion of how this renormalization might be achieved see
\cite{Ta68}.)

Furthermore, in order to simplify matters as much as possible, we
consider only distinguishable particles. Equations for
indistinguishable particles may then either be obtained by
symmetrizing or anti-symmetrizing the equations for distinguishable
particles in the usual way, or by making the necessary changes to the
Taylor method in order for it to apply to indistinguishable particles.
Taylor himself pursued the latter approach in his original
work~\cite{Ta63}. For examples of the former approach see
Ref.~\cite{Ta66} or the papers~\cite{AB80,AB81}.

The classification-of-diagrams technique is then based on the
following definitions, which apply to any perturbation diagram,
regardless of the perturbation scheme used to construct the diagram.
(Note that the definitions would have to be suitably modified if we
intended to consider the structure of the amplitude in any channel
other than the $s$-channel.)

\vskip 5 mm

\noindent {\bf Definition (r-cut)} {\em  An r-cut is an arc which
separates initial from final states and intersects exactly
$r$ lines, at least one of which must be an internal line. If all of
the $r$ lines cut are internal lines then the cut is called an
internal r-cut.}

\vskip 5 mm

Note that in writing this definition we assume that all perturbation
diagrams are represented in a two-dimensional plane. We do allow the
lines in any diagram to ``jump over" one another: two lines  do not
have to meet at an interaction vertex whenever they intersect. By
contrast, a cut is defined to intersect all the lines it meets: it may
not jump over any of them. (Other definitions of an
$r$-cut, which do not assume that the diagrams are
represented in the plane, may be composed but it is the above definition which
Taylor himself used.)

\vskip 5 mm

\noindent {\bf Definition (r-particle irreducibility)} {\em A diagram
is  called $r$-particle irreducible if, for all integers $0 \leq k
\leq r$, no $k$-cut may be made on it. An amplitude is called
$r$-particle irreducible if all diagrams contributing to it are
$r$-particle irreducible.}

\vskip 5 mm

Using these two definitions any diagram contributing to the connected
$(r-1)$-particle irreducible $m \rightarrow n$ amplitude,
$A^{(r-1) (c)}_{n \leftarrow m}$, may be placed in one of five
classes. The class of the diagram is determined by what $r$-cuts may
be made on it. The $r$-cuts which may be made on this particular
diagram are first divided as follows: if an $r$-cut is not internal it
is called ``initial" if it intersects at least one initial-state but
no final-state line; ``final" if it intersects at least one
final-state, but no initial-state line, and ``mixed" if it intersects
both initial and final-state lines. The criteria for placing the
diagram in one of the classes $C_1$--$C_5$ may now be stated as
follows:

\newcounter{class}
\begin {list}%
{$C_{\arabic{class}}$:}{\usecounter {class}
\setlength{\rightmargin}{\leftmargin}}
\item No $r$-cut may be made on the diagram, i.e. it is $r$-particle
irreducible;
\item At least one internal $r$-cut may be made on the diagram, but
no mixed or final $r$-cut is possible;
\item Only initial $r$-cuts are possible;
\item At least one mixed $r$-cut may be made, but no final $r$-cut
is possible;
\item At least one final $r$-cut may be made.
\end {list}

The process of choosing which class to place a diagram in is represented by the
flowchart in Fig.~\ref{Fig.5}. This flow chart makes it clear that any
perturbation diagram must belong to one and only one class. Therefore, we
may sum
each of
$C_1$ to $C_5$ separately, and then express $A^{(r-1) (c)}_{n
\leftarrow m}$ as the sum of the five expressions we thereby obtain.

Now, while class $C_1$ may be summed directly, the classes $C_2$--$C_5$ must
each be summed by exhibiting a unique latest $r$-cut in each diagram and so
splitting the diagram into an $r$-particle irreducible part and an
$(r-1)$-particle irreducible part. This is done via the following lemma, known
as the Last Internal Cut Lemma (LICL).

\vskip 5 mm

\noindent {\bf Lemma (Last Internal Cut)}  {\em Any $(r-1)$-particle
irreducible diagram which admits an internal $r$-cut has a unique
internal $r$-cut which is nearest to the final state.}

\vskip 5 mm

We now rehearse Taylor's proof of this result, since the structure of
the proof will turn out to be important in understanding the
double-counting problem. The proof is based on that given by Taylor
\cite{Ta63}, but has been slightly modified in order to (we hope!)\
make its structure clearer.

{\bf Proof:} Consider any two internal $r$-cuts $c_1$ and $c_2$. We
wish to find an internal $r$-cut as late or later than both of them.
If the cuts do not intersect it is clear which of the two is earlier
and which later, and the later of the two cuts is thus the internal
$r$-cut we are looking for. If they do intersect we define $c_1^-$ and
$c_2^-$ to be the portions of the two cuts nearest the initial state
and $c_1^+$ and $c_2^+$ to be the portions of the two cuts nearest the
final state. Even if the cuts intersect each other more than once
we may still proceed in this way. If an odd number of intersections occurs then
all but the first and last intersection are ignored; $c_1^+$ and $c_2^+$
($c_1^-$ and $c_2^-$) are defined to be the portions of $c_1$ and $c_2$ which
are later (earlier) than the last (first) intersection. If an even number of
intersections occur the portions of the two cuts between any two intersections
are compared individually and the resulting sequence of pieces of cut joined to
form
$c_1^+$, $c_2^+$, $c_1^-$ and $c_2^-$\footnote{Since diagrams are
two-dimensional objects these definitions are unambiguous.}. We then construct
$c^-=c_1^- \cup c_2^-$ and $c^+=c_1^+ \cup c_2^+$\footnote{Note that the use of
set notation here corresponds to viewing these cuts as sets whose members are
the lines they cut.}. These definitions of $c^-$ and
$c^+$ do not, however, tell us in which of the two sets to place a line that is
cut by {\em both} $c_1$ and $c_2$. In order to resolve this ambiguity we
proceed
as follows. The diagram under consideration may be distorted so that any line
which is cut by both $c_1$ and $c_2$ is either intersected by both cuts while
it is horizontal, or intersected by both cuts when it is vertical.
Lines which fall into the first category are called horizontal in
$c_1 \cap c_2$, and lines which fall into the second category are
called vertical in $c_1 \cap c_2$. The sets $c^-$ and
$c^+$ are then defined to both contain any line which is
horizontal in $c_1 \cap c_2$, and to both not contain any line which is
vertical in $c_1 \cap c_2$. Fig.~\ref{Fig.6} provides a pictorial
example of these definitions. Note that it is necessary to define $c^+$ and
$c^-$ in this way in order that they are completely separated from one another,
with $c^+$ nearer to the final state than $c^-$.

Now, clearly $c^+$ is nearer to the final state than either $c_1$ or $c_2$. It
is also clearly an internal cut, since it is composed entirely of internal
lines. But, is it an $r$-cut? Denote by
$N(c)$ the number of lines cut by an arc $c$. Then:
\begin{equation}
N(c_1)=r; \quad N(c_2)=r\ .                           \label {eq:rcuts}
\end{equation}
Furthermore, since the diagram in question is $(r-1)$-particle irreducible and
$c^+$ and $c^-$ both constitute cuts on it, we have:
\begin{equation}
N(c^-) \geq r; \quad \ N(c^+) \geq r\ .               \label {eq:limits}
\end{equation}
But, because of the way $c^-$ and $c^+$ are defined:
\begin{equation}
c_1 \cup c_2 \supseteq c^- \cup c^+\ ;                \label {eq:union}
\end{equation}
and
\begin{equation}
 c_1 \cap c_2 \supseteq c^- \cap c^+\ .                \label{eq:inter}
\end{equation}
Now, using:
\begin{equation}
N(A \cup B)=N(A)+N(B)-N(A \cap B)
\end{equation}
in Eq.~(\ref{eq:union}), and applying Eq.~(\ref{eq:rcuts}) gives:
\begin{equation}
2r - N(c_1 \cap c_2) \geq N(c^-) + N(c^+) - N(c^- \cap c^+)\ .
                                                      \label {eq:c1c2no}
\end{equation}
But Eq.~(\ref{eq:inter}) implies that:
\begin {equation}
N(c_1 \cap c_2) \geq N(c^- \cap c^+)\ .              \label {eq:interno}
\end {equation}
The only way Eqs.~(\ref{eq:c1c2no}), (\ref{eq:interno}) and
(\ref{eq:limits}) can be reconciled is if:
\begin {equation}
N(c^-)=N(c^+)=r\ .                                   \label{eq:liclproved}
\end {equation}
Thus $c^-$ and $c^+$ are both internal $r$-cuts, and so we have achieved our
aim
of finding an internal $r$-cut as late or later than both $c_1$ and $c_2$. Note
that (\ref{eq:liclproved}) taken with (\ref{eq:c1c2no}) and
(\ref{eq:interno}) implies that:
\begin {equation}
N(c_1 \cap c_2)=N(c^- \cap c^+)\ .
\end {equation}
When combined with (\ref{eq:inter}) this gives:
\begin {equation}
c_1 \cap c_2=c^- \cap c^+\ .
\end {equation}
Since $c^-$ and $c^+$ do not contain any line which is vertical in
$c_1 \cap c_2$, it follows that no line which is cut by $c_1$ and
$c_2$ can be vertical in $c_1 \cap c_2$.

Applying the above procedure many times allows the construction of a
unique latest $r$-cut, which is nearest to the final state. Note that
this last cut lemma apples only to {\em internal} $r$-cuts on
($r-1$)-{\em particle irreducible} diagrams, which is why we must be
careful to distinguish between class $C_2$, in which the latest
$r$-cut will obviously be an internal $r$-cut, and classes $C_3$,
$C_4$ and $C_5$, in which an $r$-cut that cuts at least one external
line may be the latest $r$-cut.

Given the above argument, it is clear that we could easily also prove
the following result:

\vskip 5 mm

\noindent {\bf Lemma (First Internal Cut)} {\em Any $(r-1)$-particle
irreducible diagram which admits an internal $r$-cut has a unique
internal $r$-cut which is nearest to the initial state.}

\vskip 5 mm

Armed with these two lemmas we now proceed to sum the five classes
$C_1$--$C_5$.

\subsection {Class $C_1$}

The sum of all $r$-particle irreducible diagrams contributing to
$A_{n \leftarrow m}^{(r-1) (c)}$ is clearly $A_{n \leftarrow m}^{(r)(c)}$, the
connected $r$-particle irreducible $m \rightarrow n$ amplitude. This,
therefore,
is the sum of class $C_1$.

\subsection {Class $C_2$}

The latest $r$-cut in any diagram in $C_2$ must be an internal
$r$-cut. Therefore, it is apparent that by using the last internal cut
lemma we may express any connected $(r-1)$-particle irreducible
diagram,
$a^{(r-1)}_{n \leftarrow m}$, which belongs to $C_2$ as:
\begin{equation}
\left[\,a^{(r)}_{n \leftarrow r}\ G^{(r)}\ a^{(r-1)}_{r
\leftarrow m}\,\right]^{(c)}\ ,                       \label{eq:C2struct}
\end{equation}
where $G^{(r)}$ is the free propagator for $r$ fully-dressed particles, and
$a^{(r-1)}_{n \leftarrow r}$ and $a^{(r)}_{r\leftarrow m}$ are two diagrams
which are ($r-1$) and $r$-particle irreducible respectively. Note that
$a^{(r-1)}_{n \leftarrow r}$ and $a^{(r)}_{r \leftarrow m}$ need not be
connected, provided that they obey the conditions discussed below. Summing over
all diagrams which contribute to $A_{n \leftarrow m}^{(r-1) (c)}$ and are in
$C_2$ then involves summing over all ($r-1$)-particle irreducible, $m
\rightarrow n$ Feynman diagrams with the structure given in
Eq.~(\ref{eq:C2struct}). Consequently, the sum of $C_2$ is:
\begin{equation}
 C_2=\left[\,A_{n \leftarrow r}^{(r)}\ G^{(r)}\ A_{r \leftarrow
m}^{(r-1)}\,\right]^{(c)}\ .                             \label {eq:C2}
\end{equation}
See Fig.~\ref{Fig.7} for a pictorial representation of this sum of
class $C_2$. The amplitudes $A_{n \leftarrow r}^{(r)}$ and
$A_{r \leftarrow m}^{(r-1)}$ may both contain disconnected pieces, as
long as:
\begin{enumerate}
\item The overall amplitude they form is connected (that is what
the superscript ${}^{(c)}$ indicates).

\item The disconnected pieces do not contain any diagrams in which a
particle (or particles) propagates freely without interacting with any
other particles.

\item The disconnected pieces of $A_{n\leftarrow r}^{(r)}$ and $A_{r \leftarrow
m}^{(r-1)}$ are such as do not allow the presence of cuts which cut less than
$r$ lines, or $r$-cuts involving final-state lines. (The presence of such cuts
is automatically forbidden if the amplitudes  $A_{n \leftarrow r}^{(r)}$ and
$A_{r \leftarrow m}^{(r-1)}$ are connected.) See, for example,
Fig.~\ref{Fig.8}, which shows a case in which a diagram apparently belonging to
the sum (\ref{eq:C2}) admits an $(r-1)$-cut. This $(r-1)$-cut is possible
because the disconnected piece of the amplitude $\A{r}{n}{r}$ may admit an
``$(r-1)$-cut" which involves only final and initial-state lines. Such a ``cut"
is not precluded by the constraint of $r$-particle irreducibility, but may
still lead to an $(r-1)$-cut when $\A{r}{n}{r}$ is used as part of a larger
diagram. Therefore we must extend the notion of $r$-particle irreducibility in
$A_{n \leftarrow r}^{(r) (d)}$ in order to prevent $l$-``cuts" with
$l \leq r$ which involve only initial and final-state lines. A
similar extension of the $(r-1)$-particle irreducibility of $A_{r
\leftarrow m}^{(r-1) (d)}$ must also be imposed. Once these revised
definitions are made the difficulty of undesirable $r$-cuts, or cuts
cutting less than $r$ lines, no longer arises here.
\end{enumerate}

\subsection {Class $C_3$}

\label {sec-Class3}

Now consider class $C_3$. We wish to take any diagram in $C_3$ and
find a unique $r$-cut nearest to the final state. However, in this case the
situation is complicated by the fact that only $r$-cuts which
intersect at least one line from the initial state are possible. It is
therefore necessary to eliminate the external lines from consideration
before applying the  last internal cut lemma.

Consider any diagram contributing to $A_{n \leftarrow m}^{(r-1)(c)}$
and in $C_3$. Construct the set of all $r$-cuts which may be made upon the
diagram. For each $r$-cut, define $r_i$ to be the number of lines from the
initial state which that cut intersects. Then, take the minimum of $r_i$ over
all possible $r$-cuts on a given diagram and denote the result by
$t_i$. This $t_i$ is then the minimum number of lines from the initial
state cut by any $r$-cut possible in this particular diagram. We call
any $r$-cut which satisfies:
\begin{equation}
r_i=t_i\ ,                                \label{eq:minimal}
\end{equation}
a minimal $r$-cut, and we denote the set of all minimal
$r$-cuts by $M_{t_i}$.  It is clear that if we can construct a unique
latest $r$-cut, $X$, out of this set
$M_{t_i}$ then no cut in any set $M_{r_i}$ will be later than
this cut $X$. In fact, the following stronger result
applies:

\vspace {5 mm}

{\bf Claim:} {\em If $(r_i + t_i) < m$ or $r \leq m$ then if the
latest cut in $M_{t_i}$ exists it must be later than all cuts in $M_{r_i}$.}

\vspace {5 mm}

{\bf Proof:} Assume a latest cut $X \in M_{t_i}$ exists. Suppose
$Y \in M_{r_i}$ is not earlier than $X$. Then $Y$ cannot be later than $X$,
therefore $X$ and $Y$ must intersect. So, construct $c^+_{XY}$ and $c^-_{XY}$
using the procedure discussed in the proof of the LICL above. That procedure
may, however, break down here since $c^-_{XY}$ need not obey $N(c^-_{XY})
\geq r$
as if $X$ and $Y$ are initial $r$-cuts, $c^-_{XY}$ may consist entirely of
initial-state lines. However, if $c^-_{XY}$ is to consist only of initial-state
lines then it must cut all of the $m$ initial-state lines. But, $c^-_{XY}$ can
cut at most $(r_i+t_i)$ initial-state lines. Therefore, if:
\begin{equation}
(r_i+t_i)<m\ ,
\end{equation}
$c^-_{XY}$ will definitely not consist only of lines from the initial-state.
Furthermore, if $r \leq m$ then the equation $N(c^-_{XY}) \geq r$ is
automatically satisfied, even if $c^-_{XY}$ is not actually a cut --- again
because if $c^-_{XY}$ is not to be a cut then it must intersect {\em all}
initial-state lines. (Actually $r<m$ leads to a contradiction, since the LICL
argument shows that if $N(c^-_{XY}) \geq r$ then $N(c^-_{XY})=r$. therefore a
$c_{XY}^-$ containing only initial-state lines is impossible for $r<m$.)

Thus if $r \leq m$ or $(r_i + t_i) < m$ the constructed $r$-cut $c_{XY}^+$ is
later than both $X$ and $Y$. But, since $c_{XY}^+$ was formed from the cuts
$X$ and $Y$ it must cut at most $t_i$ initial-state lines. Thus, either $X$
is not
the latest cut in $M_{t_i}$ or $t_i$ is not the minimum number of initial lines
cut by an $r$-cut on the diagram. Either possibility contradicts our
assumptions.
It follows that $Y \in M_{r_i}$ must be earlier than $X$.

It is obviously still necessary to construct a latest cut in $M_{t_i}$.
However,
to construct such a cut is difficult because the cuts in the set $M_{t_i}$ must
be divided as follows. Although all cuts in $M_{t_i}$ must cut the same {\em
number} of lines from the initial state they do not necessarily cut the
same {\em
set} of initial lines. Different $r$-cuts within $M_{t_i}$ may cut
different sets
of external lines $\tilde{t}_i$, as long as each such set $\tilde{t}_i$ has
$t_i$
members.  Therefore, the minimal $r$-cuts must themselves be divided
into subsets according to which group of initial-state lines $\tilde{t}_i$ they
cut. To this end we construct subsets of $M_{t_i}$, $M_{\tilde{t}_i}$, with
each
$r$-cut in the subset $M_{\tilde{t}_i}$ intersecting a specific set of
lines from
the initial state, $\tilde{t}_i$. Note that each subset $M_{\tilde{t}_i}$ still
may contain many $r$-cuts. However, within any such subset $M_{\tilde{t}_i}$
there is always a unique latest $r$-cut, constructed as follows. Consider any
$r$-cut in $M_{\tilde{t}_i}$, and suppose the lines it intersects form a
set $s$.
Now remove the lines $\tilde{t}_i$ from each set $s$ in $M_{\tilde{t}_i}$. This
turns all the cuts in $M_{\tilde{t}_i}$ into {\em internal} ($r-t_i$) cuts in
what may now be regarded as an ($r-t_i-1$)-particle irreducible diagram.
Consequently, by the last internal cut lemma, there exists a unique last
internal
cut, which cuts ($r-t_i$) lines. By joining the set of lines $\tilde{t}_i$ to
this cut we obtain a unique latest $r$-cut out of all the cuts in
this particular subset $M_{\tilde{t}_i}$.

However, in principle there are many sets $\tilde{t}_i$ and so many different
``latest" $r$-cuts will be obtained when the above procedure is applied to the
various subsets $M_{\tilde{t}_i}$. It is not immediately clear whether it is
possible to construct an {\em overall} latest minimal $r$-cut from all these
different ``latest" minimal $r$-cuts. Again, one might think that two latest
minimal cuts from two different subsets of $M_{t_i}$ could be taken and a cut
later than either of them constructed using the procedure outlined in the proof
of the last internal cut lemma. As above though, the problem is that $c^-$ may
consist entirely of lines from the initial state, and so may not be a true
cut at
all. However, repeating the proof given above but with $r_i$ set equal to $t_i$
it is clear that if $r \leq m$ or $2 t_i < m$ it will definitely be possible to
construct a unique latest cut out of the set $M_{t_i}$. Since $t_i \leq r_i
\leq
(r-1)$ it follows that if:
\begin{equation}
2(r-1) < m \quad \mbox { or } \quad r \leq m\ ,  \label {eq:uniquecondn}
\end{equation}
is satisfied then a unique latest $r$-cut in the set $M_{t_i}$ exists and
is later
than all the other cuts possible on the diagram. Since $m$ and $r$ are always
positive integers it follows that if:
\begin{equation}
r \leq m                                          \label{eq:uc1}
\end {equation}
then a unique latest $r$-cut can be found in each $C_3$ diagram contributing to
$A_{n \leftarrow m}^{(r)}$. We now have two possibilities:
\begin{enumerate}
\item $r \leq m$, in which case each diagram may be split, in a unique
fashion, into an $r$-particle irreducible and an
$(r-1)$-particle irreducible part, and so a sum for $C_3$ may be constructed.

\item $r > m$, in which case we must be more careful, since a unique
latest cut cannot be constructed directly.
\end {enumerate}

{\bf Case 1 ($r \leq m$):} Suppose that $r \leq m$ and consider any diagram
$a_{n
\leftarrow m}^{(r-1)}$ which belongs in $C_3$. For this diagram we may
construct
the set of minimal $r$-cuts $M_{t_i}$, each of which cuts precisely $t_i$
external lines. The above argument then guarantees the existence of a
unique last
$r$-cut among all those cuts in $M_{t_i}$, which is also the overall unique
latest
$r$-cut. This cut will cut a certain set of initial lines $\tilde{t}_i$.
Applying
the procedure described above, of first removing the lines $\tilde{t}_i$ from
consideration and then applying the last internal cut lemma, we find the
diagram
$a_{n \leftarrow m}^{(r-1)}$ may be expressed uniquely as:
\begin{equation}
\left[\,a_{n \leftarrow r}^{(r)}\ G^{(r/\tilde{t}_i)}\
a_{(r-t_i)\leftarrow (m-t_i)}^{(r-t_i-1)}\,\right]^{(c)} \ ,  \label{eq:C3diag}
\end{equation}
where $G^{(r/\tilde{t}_i)}$ is the free propagator for $r$ fully-dressed
particles, but with the particles in the set $\tilde{t}_i$ removed, and the
amplitudes $a_{(r-t_i) \leftarrow (m-t_i)}^{(r-t_i-1)}$ and $\asm{r}{n}{r}$
may be
disconnected, as long as these disconnected pieces are, respectively,
$r$-particle irreducible and $(r-t_i-1)$-particle irreducible in the extended
sense discussed above.

We now sum over all diagrams in $C_3$ contributing to
$A_{n \leftarrow m}^{(r-1) (c)}$, but, for the present, restrict the sum to
those
diagrams for which the minimum number of lines from the initial state which are
cut by any possible $r$-cut is $t_i$. Since this procedure involves a sum over
all topologically distinct diagrams of the form (\ref{eq:C3diag}), it follows
that the result is:
\begin{equation}
C_3^{t_i}=\sum_{\rm{All} \ {sets} \ \tilde{t}_i}
\left[\, A_{n \leftarrow r}^{(r)}\ G^{(r/\tilde{t}_i)}\
A_{(r-t_i) \leftarrow (m-t_i)}^{(r-t_i-1)}\,\right]^{(c)}\ ,\label {eq:C3ti}
\end{equation}
where, as above, the amplitudes $A_{n \leftarrow r}^{(r)}$ and $A_{(r-t_i)
\leftarrow (m-t_i)}^{(r-t_i-1)}$ may contain disconnected pieces,
provided that:
\begin {enumerate}
\item These pieces only represent processes in which each particle
interacts with
at least one other particle.

\item The irreducibility of the disconnected pieces of these amplitudes is
understood in the extended sense discussed above.

\item The overall amplitude formed in expression (\ref{eq:C3ti}) is connected.
\end {enumerate}

We have denoted the sum of this sub-class of $C_3$, which includes all
diagrams in
which the minimum number of lines from the initial state cut by any possible
$r$-cut is $t_i$, by $C_3^{t_i}$. Clearly then, the sum in Eq.~(\ref{eq:C3ti})
must be restricted to those sets $\tilde{t}_i$ with $t_i$ members. Note
also that
terms in the sum containing any one-to-one amplitude must always be eliminated,
since, due to all particles being fully dressed, we set all one-to-one
amplitudes to zero.

Before continuing we observe that, for every diagram in $C_3^{t_i}$, the cut
$\alpha$ shown in Fig.~\ref{Fig.9} must be at least an $(r+1)$-cut,
since if it is an $r$-cut the diagram belongs in $C_4$, and if it cuts
less than $r$ lines then the diagram is not $(r-1)$-particle irreducible, and
so
cannot contribute to $A_{n \leftarrow m}^{(r-1)}$. It follows that the
amplitude $A_{n \leftarrow r}^{(r)}$ in Eq.~(\ref{eq:C3ti}) must be
$(r-m+t_i-1)$-particle irreducible in the channels:
\[
[n/\tilde{h}] + [r/\tilde{t}_i] \leftarrow [\tilde{h}] + [\tilde{t}_i]\ ,
\]
where $\tilde{h}$ is any single-member set of final-state lines. It turns
out that
this condition is sufficient to stop cuts such as $\alpha$ in Fig.~\ref{Fig.9}
cutting $r$ or less lines.  If the condition is not imposed then certain
diagrams
will be included in both $C_3$ and $C_4$ and so will be double-counted. Taylor
does not seem to have realized that if this inter-class double-counting is to
be
avoided, a constraint must be placed on the structure of the amplitude $A_{n
\leftarrow r}^{(r)}$ in a channel other than the $s$-channel. The necessity of
this constraint is, in fact, merely a consequence of the fact that in
time-dependent perturbation theory, the cut-structure of the factor amplitudes
(such as $A_{n
\leftarrow r}^{(r)}$) in channels other than the $s$-channel contributes to the
overall $s$-channel cut-structure of the diagram. Observe that if $r \leq
m$ then
this constraint is automatically satisfied, due to the $s$-channel $r$-particle
irreducibility of $\A{r}{n}{r}$. Hence in the case we are discussing here these
constraints are unnecessary. However, they will become necessary if $r > m$.

Another constraint must also be imposed on $A^{(r)}_{n\leftarrow r}$ as
follows.
If $2t_i>m$, diagrams in which the cut $\beta$ shown in Fig.~\ref{Fig.9} is an
$r$-cut should have been included in $C^{m-t_i}_3$ not $C^{t_i}_3$. Therefore,
if $2t_i > m$, the condition of $(r-m+t_i)$-particle irreducibility must be
imposed on
$A^{(r)}_{n\leftarrow r}$ in the channel
\be
[n] + [r/\tilde{t}_i]\leftarrow \tilde{t}_i\ .             \label{eq:chann1}
\ee
(Note that if this condition is imposed, the condition discussed in the
previous
paragraph is automatically satisfied.) If this condition is not applied then
certain diagrams will be included in both $C^{m-t_i}_3$ and $C^{t_i}_3$ and so
inter-sub-class double-counting will occur. Furthermore, even if $2t_i\leq
m$ the
condition of $(r-m+t_i-1)$-particle irreducibility must be imposed in this
channel, since otherwise diagrams which are not $(r-1)$-particle irreducible
are
included in the sum of $C^{t_i}_3$. Thus regardless of the relative value of
$2t_i$ and $m$, the amplitude $A^{(r)}_{n\leftarrow r}$ acquires a constraint
in
the channel (\ref{eq:chann1}). Again, this constraint is automatically
satisfied
if $r\leq m$, due to the $s$-channel $r$-particle irreducibility of
$A^{(r)}_{n\leftarrow r}$.

Now, diagrams in $C_3$ contributing to $A_{n \leftarrow m}^{(r)}$ can have any
number, $t_i$, of initial lines cut by the $r$-cut in question, from a
minimum of
$t_i=1$ up to a maximum of $t_i=(r-1)$. Therefore, it follows that if condition
(\ref{eq:uniquecondn}) is satisfied:
\begin{equation}
C_3=\sum_{t_i=1}^{r-1} C_3^{t_i}\ ,               \label{eq:C3sum1}
\end{equation}
with $C_3^{t_i}$ being given by Eq.(\ref{eq:C3ti}).

{\bf Case 2 ($r > m$):} Our earlier discussion made it clear that in this
case the
existence of a unique latest $r$-cut could not be guaranteed. However, even if
$r>m$ the sub-class $C_3^{t_i}$ may still be summed, provided that $2t_i < m$ ,
as follows. If $2t_i < m$ then a unique latest $r$-cut in the set $M_{t_i}$ may
be found in each diagram in $C_3^{t_i}$. This cut may not be the unique latest
$r$-cut in the diagram but it is uniquely defined, and no later $r$-cut is
possible. Consequently this $r$-cut provides an unambiguous way of
splitting each
diagram in $C_3^{t_i}$ into two halves. When we sum over all diagrams in
$C_3^{t_i}$ we obtain the result:
\begin{equation}
C_3^{t_i}=\sum_{\rm{All} \ {sets} \ \tilde{t}_i}
\left[\,{A_{n \leftarrow r}^{(r)}}_{\tilde{t}_i}\ G^{(r/\tilde{t}_i)}
\ A_{(r-t_i) \leftarrow (m-t_i)}^{(r-t_i-1)}\,\right]^{(c)}\ , \label{eq:C3ti2}
\end{equation}
where we have had to impose on $\A{r}{n}{r}$ the restrictions discussed above,
since they are no longer automatically satisfied. Note that the sum here is
only
over those minimal sets $\tilde{t}_i$ with
$t_i$ members.

Now, if $2t_i \geq m$ and $r > m$ then we must find a different way to sum
$C_3^{t_i}$, since no unique latest $r$-cut in $M_{t_i}$ exists. Taylor claims
that in this case $C_3^{t_i}$ may be summed by splitting it into
sub-sub-classes $C_3^{\tilde {t_i}}$, where $C_3^{\tilde t_i}$ is defined to be
the set of all diagrams belonging to class $C_3$ in which a minimal $r$-cut
cutting the set of initial lines $\tilde{t}_i$ is possible. So, consider any
diagram in $C_3^{\tilde{t}_i}$. As we saw above, once a diagram and a
minimal set of lines
$\tilde{t}_i$ is chosen, there exists a unique latest $r$-cut in that
diagram, which intersects the
set of lines $\tilde{t}_i$. If the procedure described above is applied, of
first
removing the external lines $\tilde{t}_i$ from consideration and then
applying the
last internal cut lemma, we find this diagram may be written exactly as in
Eq.~(\ref{eq:C3diag}). Consequently, when we sum over all diagrams in
$C_3^{\tilde{t}_i}$ we obtain:
\begin{equation}
C_3^{\tilde{t}_i}=\left[\,{A_{n \leftarrow r}^{(r)}}_{\tilde{t}_i}
G^{(r/\tilde{t}_i)}
A_{(r-t_i) \leftarrow (m-t_i)}^{(r-t_i-1)}\,\right]^{(c)}\ , \label{eq:C3ti3}
\end{equation}
where the subscript $\tilde{t}_i$ again indicates that the conditions discussed
above have been imposed on $\A{r}{n}{r}$, in order to prevent the undesirable
cuts which would otherwise be possible.

Taylor claims that by summing over all possible minimal sets
$\tilde{t}_i$ one obtains the sum of all diagrams in $C_3$. We shall
see in Section~\ref{sec-Taylorwrong} that this mistaken claim is precisely the
origin of the double-counting problem mentioned in the Introduction. However,
if we, for the present, continue on the basis of this assumption, we find that
Eq.~(\ref{eq:C3ti2}) still holds. We may make the identification of the sum in
Eq.~(\ref{eq:C3ti2}) with a sum over all possible sets of initial lines with
$t_i$ members. This is possible because we are summing over all
topologically distinct diagrams and considering distinguishable particles.
Consequently, if the contribution from one set of initial lines $\tilde{t}_i$
is
included, the contribution from all other possible sets of initial lines with
$t_i$ members must also be included. This identification shows that, {\em given
the assumption}
$C_3^{t_i} \qeq \sum C_3^{\tilde{t}_i}$ (where the sum is defined to run over
all minimal sets $\tilde{t}_i$ with $t_i$ members), the sum of $C_3^{t_i}$ is
exactly the same in the case $r > m$ as in the case $r \leq m$, but with
restrictions imposed on $\A{r}{n}{r}$ in channels other than the $s$-channel.

If $r > m$, $t_i$ may take on values from $1$ to $(m-1)$. Therefore, it follows
that, if condition (\ref{eq:uniquecondn}) is not satisfied, the sum of class
$C_3$ may be written as:
\begin{equation}
C_3=\sum_{t_i=1}^{m-1} C_3^{t_i}\ ,                    \label{eq:C3sum2}
\end{equation}
where the sum of $C_3^{t_i}$ is still given by Eq.~(\ref{eq:C3ti2}).

The sum of $C_3$ in both of the above cases is represented diagrammatically in
Fig.~\ref{Fig.10}. Even though we have not been able to find a single unique
latest cut in all diagrams in $C_3$, the argument given above appears to
show that
we may express $C_3$ as the sum of a number of terms, in each of which
there is a
different unique latest cut.

\subsection {Class $C_4$}

\label {sec-Class4}

The argument used above to construct the sum of class $C_3$ is very similar to
that used to find the sum of $C_4$. Consider any diagram in $C_4$, and consider
any $r$-cut which can be made on that diagram. For this $r$-cut, $r_i$ is
defined
as above, and $r_f$ is defined to be the number of lines from the final state
which the cut intersects. Then, once more,  $t_i$ is defined to be the
minimum of
$r_i$, with the minimum taken over all possible $r$-cuts, and $s_f$ is
defined to
be the maximum of $r_f$, with the maximum taken over all $r$-cuts satisfying
$r_i=t_i$. This defines a set of $r$-cuts, known as minimal/maximal $r$-cuts,
$M_{s_f t_i}$ which all obey the condition:
\begin{equation}
 r_i=t_i \quad \mbox { and } \quad r_f=s_f.
\end{equation}

{\bf Claim:} {\em If there is a latest $r$-cut in the set of
minimal/maximal
$r$-cuts then this cut is later than any $r$-cut cutting $r_i > t_i$
initial-state lines and $r_f < s_f$ final-state lines, provided that:
\begin{equation}
(r_i + t_i) < m \quad \mbox { or } \quad r \leq m
\end{equation}
and
\begin{equation}
  (r_f + s_f) < n \quad \mbox { or } \quad
2r < n + \min\{m,r\} \ .
\end{equation}}

{\bf Proof:} As above, if the latest minimal/maximal $r$-cut $X$ is not already
later than the cut $Y$, which cuts $r_i$ initial-state and $r_f$ final-state
lines, then we attempt to construct a cut $c^+_{XY}$ later than both. The LICL
procedure allows us to do this, provided that:
\begin {enumerate}
\item $c_{XY}^-$ does not consist solely of lines from the initial state;

\item $c^+_{XY}$ does not consists solely of lines from the final-state,
i.e.~it is not really a cut at all.
\end {enumerate}

The condition preventing possibility 1 from occurring is:
\be
(r_i + t_i) < m\ .
\ee
As above, possibility 1 may occur without invalidating the condition
$N(c^-_{XY})\geq r$, provided that $r \leq m$.

Clearly, possibility 2 cannot occur if $(r_f + s_f) < n$. Another condition
under
which possibility 2 is forbidden can be derived, as follows.  We begin by
replacing the equation $N(c^-_{XY}) \geq r$, used in the proof of the last
internal cut lemma, by:
\begin{equation}
N(c^-_{XY})  \geq \left \{ \begin{array}{ll}
m & \mbox{ if $r > m$}\\
 r & \mbox { if $r \leq m$}
\end{array}\ , \right.
\end{equation}
since $c^-_{XY}$ may now consist entirely of lines from the initial state.
Combining this result with Eqs.~(\ref{eq:c1c2no}) and (\ref{eq:interno}) then
gives:
\begin{equation}
N(c^+_{XY}) \leq 2r - \min \{m,r\}\ .
\end{equation}
It follows that if $c^+_{XY}$ is going to consist entirely of lines from the
final state, and so invalidate the use of the last internal cut lemma argument,
the condition:
\begin{equation}
n \leq 2r - \min \{m,r\} \ ,                      \label{eq:uniquecondn22}
\end{equation}
must be satisfied. Thus possibility 2 cannot arise if:
\be
(r_f + s_f) < n \quad \mbox { or } \quad n + \min \{m,r\} >
2r.
\ee
This proves the claim.

Since $r_f,r_i \leq (r-1)$ and $s_f,t_i \geq 1$, it follows that if
the condition:
\begin{equation}
r \leq m \quad \mbox { and } \quad r < n           \label{eq:uniquecondn2}
\end{equation}
is satisfied then the latest $r$-cut in the set of minimal/maximal $r$-cuts
will
be later than any other $r$-cut possible on the diagram. Note that if $n=r$
then
the constructed latest ``$r$-cut" will merely be the set of final-state lines,
therefore we require $r<n$, in which case we find that no ``cut", $c^+_{XY}$,
consisting entirely of final-state lines can be formed from two
$r$-cuts $X$ and $Y$.

Once again, we now seek the latest $r$-cut within the set of minimal/maximal
$r$-cuts. This may be done, using the LICL procedure, provided that the
conditions 1 and 2 discussed above are met. As was seen in the previous
subsection, condition 1 leads to the requirement:
\begin{eqnarray}
2t_i < m \quad \mbox { or } \quad r \leq m\ .
\end {eqnarray}
Condition 2 will automatically be satisfied if:
\be
2s_f < n\ .
\ee
Furthermore, as above, condition 2 will also be satisfied, provided that
Eq.~(\ref{eq:uniquecondn22}) is obeyed. Consequently the condition under which
the second of the two above possibilities becomes forbidden is:
\begin{equation}
2s_f<n \quad \mbox { or } \quad 2r <n + \min \{m,r\} \ .
\end{equation}
Since $t_i$ and $s_f$ are both less than or equal to $(r-1)$ it follows
that the condition which guarantees that there is a unique latest $r$-cut for
each diagram in $C_4$ is merely Eq.~(\ref{eq:uniquecondn2}).

Therefore the argument used in the proof of the last internal cut lemma may
definitely be used to construct a unique latest cut out of all the
minimal/maximal $r$-cuts, and this cut will also be the unqiue latest $r$-cut
of
all cuts in this diagram if  Eq.~(\ref{eq:uniquecondn2}) is satisfied. In this
case it is guaranteed that each diagram contributing to $\A{m}{n}{r}$ will
have a unique latest $r$-cut. Condition (\ref{eq:uniquecondn2}) is, in fact, a
less stringent condition for the success of the last internal cut lemma
argument than the condition which was used by Taylor:
\begin{equation}
2(r-1) < n,m\ .
\end{equation}

We now consider two cases:

{\bf Case 1 ($r \leq m$ and $r < n$):} The existence of a unique latest $r$-cut
allows, via the use of techniques similar to those used to sum the sub-class
$C_3^{t_i}$, the summation of all diagrams in the sub-class $C_4^{s_f t_i}$.
The
sub-class $C_4^{s_f t_i}$ is defined to contain all diagrams in $C_4$ in which
the minimal/maximal $r$-cut intersects $s_f$ lines from the final state and
$t_i$ lines from the initial state. Its sum is:
\begin{equation}
C_4^{s_f t_i}=\sum_{\rm{All} \ {sets} \ \tilde{s}_f \
\rm{\&} \ \tilde{t}_i}
\left[\,A_{(n-s_f) \leftarrow (r-s_f)}^{(r-s_f)}\
G^{(r/(\tilde{s}_f \cup \tilde{t}_i))}\
A_{(r-t_i) \leftarrow (m-t_i)}^{(r-t_i-1)}\,\right]^{(c)}\ , \label{eq:C41}
\end{equation}
where the sum is constrained to be only over those sets $\tilde{s}_f$ and
$\tilde{t}_i$ containing $s_f$ and $t_i$ members respectively, and all
one-to-one amplitudes are to be set to zero. Here,
$G^{(r/(\tilde{s}_f \cup \tilde{t}_i))}$ is the free propagator for $r$
fully-dressed particles, but with any particle which is in either of the
two sets
$\tilde{t_i}$ and $\tilde{s_f}$ removed. It follows that the sum of $C_4$ can
be
constructed by summing over all possible values of $s_f$ and $t_i$, yielding:
\begin{equation}
C_4=\sum_{s_f=1}^{r-1} \sum_{t_i=1}^{r-1} C_4^{s_f t_i}\ ,\label {eq:case1res}
\end{equation}
where the sum over $s_f$ and $t_i$ is restricted to those $s_f$ and $t_i$ which
obey:
\begin{equation}
(s_f + t_i) \leq (r-1)\ .
\end{equation}

{\bf Case 2 ($r \geq n$ or $r >m$):} On the other hand, if condition
(\ref{eq:uniquecondn2}) is not satisfied, we may sum classes $C_4^{s_f t_i}$
which obey $2s_f < n$ and $2t_i < m$, by exhibiting the unique latest cut in
the
set of minimal/maximal $r$-cuts. Again, this cut is not the unique latest cut
on
the diagram, but there is no later cut, and the cut is uniquely defined. We
find
that if $r\geq n$ or  $r<m$ but $2s_f < n$ and $2t_i < m$ then $C_4^{s_f t_i}$
is still given by Eq.~(\ref{eq:C41}), subject to certain restrictions
discussed below.

Furthermore, if $2s_f \geq n$ and $2t_i \geq m$, and $2s_f-n+2t_i-m > 0$, then
the cut indicated in Fig.~\ref{Fig.11}, which may be made on any diagram in
$C_4^{s_f t_i}$, is an $l$-cut, with $l<r$. Therefore it follows that we can
never have:
\be
(2s_f-n+2t_i-m )> 0\ ,
\ee
or otherwise an $l$-cut with $l<r$ will be possible on every diagram in
$C_4^{s_f t_i}$. Consequently  if $2s_f \geq n$ and $2t_i \geq m$ the
situations:
\begin{equation}
2s_f>n \quad \mbox{ and } \quad 2t_i \geq m
\end{equation}
or
\begin{equation}
2s_f \geq n \quad \mbox{ and } \quad 2t_i > m\ ,
\end{equation}
are forbidden, and only $(2s_f - n + 2t_i -m )=0$ is allowed.

If $2s_f \geq n$ or $2t_i \geq m$, while $(2s_f-n+2t_i-m) \leq 0$ Taylor claims
that we may still sum $C_4$ by splitting it into sub-sub-classes
$C_4^{\tilde{s}_f \tilde{t}_i}$. Here $C_4^{\tilde{s}_f \tilde{t}_i}$ is
defined
to be the set of all diagrams belonging to $C_4$ for which some minimal/maximal
$r$-cut intersects the lines $\tilde{t}_i$ from the initial state and the lines
$\tilde{s}_f$ from the final state. Once again, similar arguments to the above
allow Taylor to show that when all contributions to
$C_4^{\tilde{s}_f \tilde{t}_i}$ are summed:
\begin{equation}
C_4^{\tilde{s}_f \tilde{t}_i}=\left[\,A_{(n-s_f) \leftarrow
(r-s_f)}^{(r-s_f)}\ G^{(r/(\tilde{s}_f \cup \tilde{t}_i))}
\ A_{(r-t_i) \leftarrow (m-t_i)}^{(r-t_i-1)}\,\right]^{(c)}\ ;
\label{eq:C4sfti}
\end{equation}
from which he obtains:
\begin{equation}
C_4^{s_f t_i} \qeq \sum_{\rm{All} \ {sets} \ \tilde{s}_f \
\rm{\&} \ \tilde{t}_i} \left[\,A_{(n-s_f) \leftarrow (r-s_f)}^{(r-s_f)}\
G^{(r/(\tilde{s}_f \cup \tilde{t}_i))}\
A_{(r-t_i) \leftarrow (m-t_i)}^{(r-t_i-1)}\,\right]^{(c)}\ , \label{eq:C4}
\end{equation}
where the sum is restricted to those sets $\tilde{s}_f$ and $\tilde{t}_i$ which
contain, respectively, $s_f$ and $t_i$ members. Again, we note that the
following facts about this result:
\begin {enumerate}
\item We question Taylor's moving from Eq.~(\ref{eq:C4sfti})
to Eq.~(\ref{eq:C4}), for the reasons to be detailed in
Section~\ref{sec-Taylorwrong}.
\item  All one-to-one amplitudes in the sum must be set to zero.
\end {enumerate}

Note that in both (\ref{eq:C41}) and (\ref{eq:C4}), the factor
amplitudes $\A{(r-s_f)}{(n-s_f)}{r-s_f}$ and $\A{(m-t_i)}{(r-t_i)}{r-t_i-1}$
may be disconnected, provided that the disconnected parts are, respectively,
$(r-s_f)$ and $(r-t_i-1)$-particle irreducible in the extended sense discussed
above.

Furthermore, in order to stop diagrams which should be in $C_5$ also being
included in $C_4$, and so being double-counted, certain restrictions must be
placed on these sub-amplitudes in channels other than the $s$-channel. In fact,
the presence of the cut
$\alpha$ shown in Fig.~\ref{Fig.12} shows that $\A{(m-t_i)}{(r-t_i)}{r-t_i-1}$
must be $(r-n+s_f)$-particle irreducible in the channel:
\[
[\tilde{s}_f]\leftarrow [(r-t_i)/\tilde{s}_f] + [m-t_i]\ .
\]
Note that this
condition is automatically satisfied if $r < n$, due to the
$s$-channel $(r-t_i-1)$-particle irreducibility of
$\A{(m-t_i)}{(r-t_i)}{r-t_i-1}$.

A similar problem arises because the cut $\beta$ shown in Fig.~\ref{Fig.12} may
be drawn. If $2t_i > m$, diagrams in which this cut is an $r$-cut should have
been placed in the sub-class $C_4^{1 \, (m-t_i)}$. Thus, if $2t_i > m$ we must
enforce the restriction of $(r-m+t_i-1)$-particle irreducibility on the
amplitude $\A{(r-s_f)}{(n-s_f)}{r-s_f}$ in all channels:
\[
[(r-s_f)/\tilde{t}_i] +  [(n-s_f)/\tilde{h}] \leftarrow [\tilde{h}] +
[\tilde{t}_i] \ ,
\]
where $\tilde{h}$ is any single-member set of final-state lines. Note that
even if
$2t_i \leq m$ the amplitude must be $(r-m+t_i-2)$-particle irreducible in this
channel, as otherwise a cut involving less than $r$ lines will be possible on
some diagrams summed in $C_4^{s_f t_i}$. Note also that if $r \leq m$ this
condition is satisfied automatically, due to the $s$-channel
$(r-s_f)$-particle irreducibility of $\A{(r-s_f)}{(n-s_f)}{r-s_f}$. If $r > m$
the presence of this restriction in the expression for $C_4^{s_f t_i}$ is
necessary, in order to ensure that we do not include any diagrams in this
sub-class which should actually have been included in other sub-classes of
$C_4^{s_f t_i}$ (and so produce inter-sub-class double-counting), or any
diagrams
which are, in fact, not $(r-1)$-particle irreducible.

Thus, instead of the expression (\ref{eq:C4})  we
must write:
\begin{equation}
C_4^{s_f t_i} \qeq \sum_{\rm{All} \ {sets} \ \tilde{s}_f \
\rm{\&} \ \tilde{t}_i} \left[\,{A_{(n-s_f) \leftarrow
(r-s_f)}^{(r-s_f)}}_{\tilde{t}_i}\ G^{(r/(\tilde{s}_f \cup
\tilde{t}_i))}\ {A_{(r-t_i) \leftarrow
(m-t_i)}^{(r-t_i-1)}}_{\tilde{s}_f}\,\right]^{(c)}\ ,     \label{eq:C42}
\end{equation}
where the subscripts $\tilde{t}_i$ and $\tilde{s}_f$ indicate that the
restrictions discussed above have been imposed. Note that modification to
Eq.~(\ref{eq:C41}) is not necessary  since if
$r<n$ and $r\leq m$ the conditions represented by the subscripts $\tilde{t}_i$
and $\tilde{s}_f$ are automatically satisfied.

The possible values of $s_f$ and $t_i$ may then be summed over in order to
yield:
\begin{equation}
C_4=\sum_{t_i=1}^{\min\{m,r\}-1} \
\sum_{s_f=1}^{\min\{n,r\}-1} C_4^{s_f t_i}\ ,           \label{eq:C4sum}
\end{equation}
where, once again, the sums are restricted to $(t_i + s_f) \leq (r-1)$ and
$(m-2t_i+n-2s_f) \geq 0$. This result in fact encompasses
Eq.~(\ref{eq:case1res}),
which applies only to the case $r \leq m$ and $r < n$. For a diagrammatic
representation of the sum of class $C_4$ see Fig.~\ref{Fig.13}.

\subsection {Class $C_5$}

The method for summing classes $C_3$ and $C_4$ is very similar to that used in
order to sum class $C_5$. Consider any diagram in class $C_5$ and consider any
particular $r$-cut which can be made on that diagram. Define $r_f$ to be the
number of lines from the final state cut by that $r$-cut. The maximum of $r_f$
over all possible $r$-cuts is taken and is defined to be $s_f$. The set of
$r$-cuts satisfying:
\begin{equation}
 r_f=s_f,
\end{equation}
is defined to be $M_{s_f}$, the set of maximal $r$-cuts. Once again, a unique
latest cut may be extracted from this set of $r$-cuts, and shown to be
later than
any $r$-cut involving $r_f$ final-state lines, $r_f < s_f$, provided that the
argument used in the proof of the last internal cut lemma is applicable. In the
previous section we explained two ways in which this argument might break down
when applied to a diagram in class $C_4$. When the argument is applied to a
diagram in class $C_5$ it cannot break down in the first of these two ways,
since, in this case, it is certain that $c^-$ does not contain any
initial-state
lines. Therefore, the only way the last internal cut lemma argument can fail
when applied to a diagram in $C_5$ is if $c^+$ contains only lines from the
final
state. Arguing as we did above shows that $c^+$ cannot contain only lines from
the final state if:
\begin{equation}
(r-1+s_f) <n  \quad \mbox { or } \quad r<n\ .           \label{eq:ucn}
\end{equation}
Since $s_f \geq 1$, the
condition for there to definitely be a unique latest cut among all the cuts in
the set $M_{s_f}$, and for that cut to be later than all other $r$-cuts
possible on the diagram, is found to be:
\begin{equation}
r<n\ .                                                 \label{eq:uc3}
\end{equation}
We note that condition (\ref{eq:uc3}) is slightly different from the condition
used by Taylor. He stated that the condition for the generation of a unique
last
cut in the set $M_{s_f}$ was $r \leq n$. However, the above discussion shows
that the argument used in the proof of the last internal cut lemma may well
also
fail to generate a unique last cut if $n=r$.

Once more, we now consider two cases:

{\bf Case 1 ($r < n$):} If condition (\ref{eq:uc3}) holds then similar
arguments
to those used above may be employed in order to show that:
\begin{equation}
C_5=\sum_{s_f=1}^{r-1} C_5^{s_f}\ ;                  \label{eq:C5sum1}
\end{equation} with:
\begin{equation}
C_5^{s_f}=\sum_{\rm{All} \ {sets} \ \tilde{s}_f}
 \left[\,A_{(n-s_f) \leftarrow (r-s_f)}^{(r-s_f)}\ G^{(r/\tilde{s}_f)}
\ A_{r \leftarrow m}^{(r-1)}\,\right]^{(c)}\ ,       \label{eq:C5sf}
\end{equation}
where the sum is restricted to those sets $\tilde{s}_f$ with
$s_f$ members, and both $\A{m}{r}{r-1}$ and $\A{(r-s_f)}{(n-s_f)}{r-s_f}$
may contain disconnected parts, subject to the restrictions discussed
above for disconnected parts.

{\bf Case 2 ($r\geq n$):} If condition (\ref{eq:uc3}) is violated,
then provided that $2s_f < n$ the sub-class $C_5^{s_f}$ may still
be summed to yield (\ref{eq:C5sf}), again subject to the corrections
discuused below. If $2s_f \geq n$ then  Taylor splits $C_5$ into
sub-sub-classes,
$C_5^{\tilde{s}_f}$, each of which contains all those diagrams which admit a
maximal $r$-cut intersecting the set of final-state lines $\tilde{s}_f$. When
$C_5^{\tilde{s}_f}$ is summed its sum is found to be:
\begin{equation}
C_5^{\tilde{s}_f}=\left[\,A_{(n-s_f) \leftarrow
(r-s_f)}^{(r-s_f)}\ G^{(r/\tilde{s}_f)}\
A_{r \leftarrow m}^{(r-1)}\,\right]^{(c)}.
\end{equation}
By summing over all possible sets $\tilde{s}_f$ with $s_f$ members Taylor
obtains
Eq.~(\ref{eq:C5sf}) for $C_5^{s_f}$, with the same comments which applied to
that result still applying here. Note that, as we did for $C_3$ and $C_4$, we
question this last step for the reasons described below. However, if this step
is accepted, summing over $s_f$ gives:
\begin{equation}
 C_5=\sum_{s_f=1}^{n-1} C_5^{s_f}\ .               \label{eq:C5sum}
\end{equation}

Once more, the $r$-cut depicted in Fig.~\ref{Fig.14} must be prohibited if
$2s_f
< n$, and the possibility of $l$-cuts, with $l < r$ must be stopped, regardless
of the value of $s_f$. Thus we impose the restriction of $(r-n+s_f)$-particle
irreducibility if $2s_f < n$, and $(r-n+s_f-1)$-particle irreducibility if
$2s_f
\geq n$, on $\A{m}{r}{r-1}$ in the channel:
\[
[\tilde{s}_f]\leftarrow[r/\tilde{s}_f] + [m]\ .
\]
Note that since the amplitude is automatically $(s_f-1)$-particle irreducible
in this channel these conditions are automatically satisfied if $r< n$.
Therefore we adjust the equation for $C_5^{s_f}$ to read:
\begin {equation}
C_5^{s_f} \qeq \sum_{\rm{All} \ {sets} \ \tilde{s}_f}
 \left[\,A_{(n-s_f) \leftarrow (r-s_f)}^{(r-s_f)}\ G^{(r/\tilde{s}_f)}\
{A_{r \leftarrow m}^{(r-1)}}_{\tilde{s_f}}\,\right]^{(c)}\ . \label{eq:C5sum2}
\end {equation}
For a diagrammatic representation of the sum of class $C_5$, see
Fig.~\ref{Fig.15}.

\subsection {Overall result}

This achieves our original aim of finding expressions for each of the classes
$C_1$ to $C_5$. If we now sum the results of our summation of each of
the individual classes we find:
\begin{eqnarray}
A_{n \leftarrow m}^{(r-1) (c)} &\qeq& A_{n \leftarrow
m}^{(r) (c)} + \left[\,A_{n \leftarrow r}^{(r)} G^{(r)} A_{r
\leftarrow m}^{(r-1)}\,\right]^{(c)} \nonumber\\
&+& \sum_{t_i=1}^{\min\{m,r\}-1} \sum_{\rm{All} \ {sets} \ \tilde{t}_i}
 \left\{ {A_{n \leftarrow r}^{(r)}}_{\tilde{t}_i} G^{(r)}
\left[{G^{\tilde{t}_i}}^{-1} A_{(r-t_i) \leftarrow
(m-t_i)}^{(r-t_i-1)} \right] \right \}^{(c)} \nonumber\\
&+& \sum_{s_f=1}^{\min\{n,r\}-1} \sum_{t_i=1}^{\min\{m,r\}-1}
\sum_{\rm{All} \ {sets} \ \tilde{s}_f \ \rm{\&} \ \tilde{t}_i}
\left \{ \left[ {A_{(n-s_f) \leftarrow
(r-s_f)}^{(r-s_f)}}_{\tilde{t}_i} {G^{\tilde{s}_f}}^{-1}
\right] G^{(r)} \right.  \nonumber \\
& & \hskip 6 cm\times\ \left. \left[{G^{\tilde{t}_i}}^{-1}
{A_{(r-t_i) \leftarrow (m-t_i)}^{(r-t_i-1)}}_{\tilde{s}_f} \right]
\right \}^{(c)}\nonumber\\
&+& \sum_{s_f=1}^{\min\{n,r\}-1}
\sum_{\rm{All}
\ {sets} \
\tilde{s}_f} \left \{ \left[A_{(n-s_f) \leftarrow (r-s_f)}^{(r-s_f)}
{G^{\tilde{s}_f}}^{-1} \right] G^{(r)}
{A_{r \leftarrow m}^{(r-1)}}_{\tilde{s}_f} \right\}^{(c)},
\label{eq:Amnfull}
\end{eqnarray}
where the superscript ${}^{(c)}$ indicates that only connected diagrams may be
formed. Note that the five terms in this equation are each generated by a
different Taylor class, with the $n$th term generated by $C_n$, where $n=1
\ldots 5$ and the question mark above the equals sign reminds us of the
queries raised about the validity of certain of the above steps in  the case
$r \geq n$ or $r > m$.

We observe that in Eq.~(\ref{eq:Amnfull}) we have expressed
$A_{n \leftarrow m}^{(r-1) (c)}$ in terms of amplitudes of equal or greater
irreducibility and amplitudes involving fewer particles. We have done this
without making any assumption about the structure of the underlying field
theory, other than the fact that the theory has a perturbation expansion in
terms of Feynman diagrams. This model-independence is what makes the Taylor
method so powerful and useful.

Indeed, as was mentioned above, the Taylor method is valid even if no
underlying
field theory exists at all. The only prerequisite for an application of the
Taylor method is the presence of a diagrammatic expansion. Therefore, the
Taylor
method may be applied to a system of $N$ particles, in order to derive
equations
for the $N$-particle amplitudes in terms of the $n$-particle amplitudes,
where $n
< N$. When used in a three-particle system such a procedure results in
Faddeev-type equations\cite{Fa61}, and when it is used in a four-particle
system
this procedure will lead to Yakubovski\u{i}-type equations\cite{Ya67}. Thus,
not
only is the Taylor method an extremely valuable model-independent technique
within field theory, but it is also applicable to other problems, e.g. those in
the theory of $N$-particle systems.

\section {The relationship of Taylor's original method to the TRAB
simplification} \label {sec-ABTaylor}

In the last section we reviewed the Taylor method and showed how it allows us
to
write the amplitude $A_{n \leftarrow m}^{(r-1)(c)}$ in terms of amplitudes of
equal or greater irreducibility and amplitudes involving fewer particles.
However, since 1979, a simpler version of the Taylor method has also been used.
This simplification was first developed for a time-ordered perturbation theory
by Thomas and Rinat \cite{Th73,TR79}. It was then applied by Afnan and
Blankleider to the $NN-\pi NN$ system \cite{AB80,AB85}, by Afnan and Pearce to
the $\pi N-\pi \pi N$ system \cite{AP86,AP87}, and by Afnan and Araki to
the problem of pion photoproduction on the nucleon and deuteron
\cite{AA87,Af88,AA88}. In this paper we refer to this simplified method as
the TRAB method, and we begin this section by reviewing the method. It
has been suggested, most notably by Avishai and Mizutani \cite{AM83}, that,
when
applied to a {\em time-dependent perturbation theory}, the TRAB method of
classifying diagrams leads to results different from those obtained using
Taylor's original method. In order to establish the exact conditions under
which
the TRAB and Taylor methods are equivalent we examine the expression obtained
for $A_{n\leftarrow m}^{(r-1)(c)}$ in the previous section and compare it to
that obtained in this section from the TRAB method.

The TRAB method is a simplification of the Taylor method which was explicitly
designed only to apply to a time-ordered perturbation theory without
anti-nucleons \cite{TR79}. In the TRAB method the definition of an $r$-cut and
an $r$-particle irreducible diagram are exactly those given for Taylor's method
in the previous section, but with the restriction that, since TRAB deal only
with time-ordered diagrams, cuts can only be vertical lines separating the
initial and final states. Once cuts are restricted to vertical lines the
last-cut
lemma is trivial to prove, and many of the restrictions imposed in Taylor's
work
in order to guarantee the existence of a unique last cut become unnecessary.
The
last-cut lemma in the TRAB method may therefore be stated as:

\vskip 5 mm

\noindent {\bf Lemma (TRAB last-cut)} {\em There exists a unique latest $r$-cut
in any time-ordered perturbation theory diagram whose irreducibility $k$ is
less
than $r$.}

\vskip 5 mm

Of course, strictly speaking, this version of the last-cut lemma is only
valid in
time-ordered perturbation theory, but this new last-cut lemma is much easier to
use than the older, more general, Taylor version. In order to use the TRAB
last-cut lemma to find an equation for the connected amplitude, $A_{n
\leftarrow
m}^{(r-1)(c)}$, we merely observe that all diagrams contributing to this
amplitude must be either $r$-particle irreducible or $r$-particle reducible.
The
sum of the diagrams in the first group is clearly the fully-connected
$r$-particle  irreducible amplitude $A_{n \leftarrow m}^{(r)(c)}$. The
sum of the diagrams in the second group is, by the last-cut lemma:
\begin{equation}
\left[\, \bar{A}_{n \leftarrow r}^{(r)}\ G^{(r)}\ \bar{A}_{r \leftarrow
m}^{(r-1)}\, \right]^{(c)}\ ,
\end{equation}
where the amplitude $\bar{A}$ may contain both connected and disconnected
pieces,
and the superscript ${}^{(c)}$ indicates that the overall diagram must be
connected. Note that the disconnected part of the amplitude $\bar{A}$ {\em may}
contain diagrams in which one or more particles merely propagate freely.
Putting the sums of the two groups of diagrams together implies that the TRAB
method gives the following equation for
$A_{n \leftarrow m}^{(r-1)(c)}$:
\begin{equation}
A_{n \leftarrow m}^{(r-1)(c)}=A_{n \leftarrow m}^{(r)(c)} +
\left[\, \bar{A}_{n \leftarrow r}^{(r)}\ G^{(r)}\ \bar{A}_{r \leftarrow
m}^{(r-1)}\, \right]^{(c)}\ .
\end{equation}
That is:
\begin{equation}
A_{n \leftarrow m}^{(r-1)(c)}=A_{n \leftarrow m}^{(r)(c)} +
\left\{ \left[\,A_{n \leftarrow r}^{(r)(c)} + \bar{A}_{n \leftarrow
r}^{(r)(d)}\, \right] G^{(r)}
\left[\, A_{r \leftarrow m}^{(r-1)(c)} + {\bar{A}_{r \leftarrow
m}^{(r-1)}}{}^{(d)}\, \right]\right\}^{(c)}\ .       \label{eq:AmnTRAB}
\end{equation}
As mentioned above, the disconnected amplitudes $\bar{A}^{(d)}$ may contain
terms
in which one or more particles merely propagate freely while the others
interact.

This technique was applied by Afnan and Blankleider to the $NN-\pi NN$ and
$BB-\pi BB$ problems in what was, apparently, a covariant approach
\cite{AB80,AB85}. This would appear to be incorrect since the TRAB approach was
originally designed to be applied only to a time-ordered perturbation theory
without anti-nucleons. The remarkable thing is that Afnan and Blankleider's
application of the TRAB technique produced exactly the same equations for the
$NN-\pi NN$ system as those obtained by Avishai and Mizutani using the full
Taylor method~\cite{Mi76,AM83}. Avishai and Mizutani suggested that this
coincidence of equations required investigation. Here we discuss this
coincidence and find that it occurs only because Avishai and Mizutani
ignored the
restrictions on amplitudes in channels other than the $s$-channel ---
restrictions which, in the previous section, we found were necessary if the
correct equation for $\A{m}{n}{r-1}$ was to be derived.

In order to establish the connection we rewrite Eq.~(\ref{eq:AmnTRAB}) using
the
definitions:
\begin{equation}
\bar{A}^{(r-1) (d)}_{r \leftarrow m}\bar{A}_{r \leftarrow m}^{(r-1)
(\tilde{d})} + \sum_{t_i=1}^{\min\{m,r\}-1}
\sum_{\rm{All} \ {sets} \ \tilde{t}_i} {G^{\tilde{t}_i}}^{-1}
\left[\,\bar{A}_{(r-t_i) \leftarrow(m-t_i)}^{(r-t_i-1) (c)}+\bar{A}_{(r-t_i)
\leftarrow (m-t_i)}^{(r-t_i-1) (\tilde{d})}\,\right]     \label{eq:dc1}
\end{equation}
and:
\begin{equation}
\bar{A}^{(r) (d)}_{n \leftarrow r}=\bar{A}_{n \leftarrow r}^{(r)
(\tilde{d})} + \sum_{s_f=1}^{\min\{n,r\}-1}
\sum_{\rm{All} \ {sets} \ \tilde{s}_f} \left[\,\bar{A}_{(n-s_f) \leftarrow
(r-s_f)}^{(r-s_f) (c)}+\bar{A}_{(n-s_f) \leftarrow
(r-s_f)}^{(r-s_f) (\tilde{d})}\,\right]\
{G^{\tilde{s}_f}}^{-1}\ ,                                 \label{eq:dc2}
\end{equation}
where in each case $\bar{A}_{f \leftarrow i}^{(I) (\tilde{d})}$ is the
disconnected piece of the $I$-particle irreducible $i \rightarrow f$ amplitude,
but with no particles propagating freely, i.e. every particle interacting with
at least one other particle. Having made these definitions it is
obvious that:
\begin {eqnarray}
\bar{A}_{(r-t_i)\leftarrow (m-t_i)}^{(r-t_i-1)
(c)} + \bar{A}_{(r-t_i) \leftarrow (m-t_i)}^{(r-t_i-1) (\tilde{d})}
&=& A_{(r-t_i) \leftarrow (m-t_i)}^{(r-t_i-1)}\ ;\quad
t_i=0,1,2,\ldots,\min\{m,r\}-1\\
\bar{A}_{(n-s_f) \leftarrow (r-s_f)}^{(r-s_f) (c)} +  \bar{A}_{(n-s_f)
\leftarrow (r-s_f)}^{(r-s_f)
(\tilde{d})} &=& A_{(n-s_f) \leftarrow (r-s_f)}^{(r-s_f)}\ ;\quad
s_f=0,1,2,\ldots,\min \{ n,r \}-1,
\end {eqnarray}
provided that we interpret the disconnected amplitude $\bar{A}_{f
\leftarrow i}^{(I) (\tilde{d})}$ to be $I$-particle irreducible in the extended
sense introduced in the previous section. It follows that
Eq.~(\ref{eq:AmnTRAB})
may be rewritten as:
\begin{eqnarray}
A_{n \leftarrow m}^{(r-1) (c)}=A_{n
\leftarrow m}^{(r) (c)}
 + \left \{ \left[ A_{n \leftarrow r}^{(r)} +
\sum_{s_f=1}^{\min\{n,r\}-1}
\sum_{\rm{All} \ {sets} \ \tilde{s}_f}
 A_{(n-s_f) \leftarrow (r-s_f)}^{(r-s_f)} {G^{\tilde{s}_f}}^{-1}
\right]
\right.\nonumber\\ G^{(r)} \left. \left[ A_{r \leftarrow m}^{(r-1)} +
\sum_{t_i=1}^{\min\{m,r\}-1} \sum_{\rm{All} \ {sets} \ \tilde{t}_i}
{G^{\tilde{t}_i}}^{-1} A_{(r-t_i) \leftarrow
(m-t_i)}^{(r-t_i-1)}\right] \right \}^{(c)}\ .   \label{eq:AmnTRABsimp}
\end{eqnarray}

This equation is to be compared to the equivalent equation obtained from
the full
Taylor method, Eq.~(\ref{eq:Amnfull}).  (Note that in using
Eq.~(\ref{eq:Amnfull}) we are completely ignoring the problem of
double-counting
between the sub-sub-classes $C^{\tilde{n}}$ of a particular sub-class $C^n$ of
some class $C$. We shall return to this difficulty in the next section.)
But, provided that the restrictions represented  by the subscripts
$\tilde{s}_f$ and $\tilde{t}_i$ are ignored, Eq.~(\ref{eq:Amnfull}) may
be simplified in order to obtain Eq.~(\ref{eq:AmnTRABsimp}).

Thus if Eq.~(\ref{eq:Amnfull}) is accepted as correct the TRAB method and
Taylor's original method produce the same result, provided that the
restrictions which were placed on $\A{m-t_i}{r-t_i}{r-t_i-1}$ for
$t_i=0,1,2,\ldots,\min\{m,r\}-1$ and
$\A{r-s_f}{n-s_f}{r-s_f}$ for $s_f=0,1,2,\ldots,\min\{n,r\}-1$ are ignored.
These restrictions were imposed in order to prevent the illegitimate $l$-cuts,
$l \leq r$, which are otherwise possible on the diagrams representing the sums
of $C_3$--$C_5$. They do not arise in the TRAB method, since that method sprung
from classifying time-ordered perturbation theory diagrams, and so the
amplitudes involved do not have their irreducibility constrained in any
channel other than the $s$-channel. Therefore, when applied in a
time-dependent perturbation theory, the TRAB method includes certain diagrams
which are actually members of $C_4$ in
$C_3$ as well, and certain diagrams which are in $C_5$ in $C_4$ as well. It
also
includes some diagrams in two sub-classes $C^n$ of a particular class $C$.
Worse still, the TRAB method's failure to produce constraints in channels other
than the
$s$-channel may mean that certain diagrams which are not
$(r-1)$-particle irreducible are included in $C_3$, $C_4$ and $C_5$. However,
if
$n,m$ and $r$ are less than or equal to three it can be shown that this final
difficulty does not arise. Furthermore, if $n,m,r \leq 3$ the TRAB method does
not lead to diagrams being included in more than one sub-class $C^n$ of the
same
class. Therefore in the case $n,m,r \leq 3$, the only problem with the TRAB
method
is that it produces expressions for $C_3$--$C_5$ which mean that:
\bea
C_3 \cap C_4 &\neq& \phi\ ;\\
C_4 \cap C_5 &\neq& \phi\ .
\eea

Now in their derivation of the $NN-\pi NN$ equations Avishai and Mizutani {\em
did} ignore this difficulty as, to some extent, did Taylor himself.
Consequently, it is not in the least surprising that Avishai and Mizutani's
application of the Taylor method produced the same $NN-\pi NN$ equations as the
TRAB method. However, if the Taylor method had been applied correctly in
Avishai and Mizutani's work, and the conditions $C_3 \cap C_4=\phi$ and
$C_4 \cap
C_5=\phi$ rigorously enforced, then they would have obtained different
equations
to those derived via the TRAB method.

Consequently, care must be exercised when applying the TRAB method to
the classification-of-diagrams in time-dependent perturbation theory. But, in
essence the only problem with the TRAB method is a general one with any method
based on Taylor's  work: all such methods contain the possibility of
double-counting, as mentioned in the Introduction. We have already seen that
double-counting between different classes or different sub-classes (as well as
other problems associated with the existence of undesirable cuts) may arise if
care is not taken in the summation of the classes. We now discuss a different
sort of double-counting, in which the sum of a particular sub-class includes
certain diagrams twice.

\section {The double-counting problem in the Taylor method}
\label{sec-Taylorwrong}

Taylor's classification-of-diagrams technique was developed in an effort to sum
the diagrammatic expansion obtained for some Green's function or  amplitude in
a
perturbation theory, counting each topologically distinct diagram once and only
once. In this section we first give two examples which show that, in a
time-dependent perturbation theory, such as covariant perturbation theory, the
method fails to achieve this aim. I.e., when the Taylor method is used to sum
certain series of time-dependent perturbation theory diagrams some diagrams in
the series are counted twice. The flaw in Taylor's argument which leads to this
double-counting is then explained, and it is shown how that mistake leads
directly to the two examples of double-counting given.

Note that the type of double-counting discussed here is fundamentally different
from that mentioned above. Even though Taylor appears not to have realized the
full extent of this first type of double-counting, he does appear to have
eliminated it in certain cases. Indeed, the inter-class and
inter-sub-class double-counting mentioned so far may be eliminated by
taking care
when summing $C_3$--$C_5$, and placing appropriate restrictions on the
amplitudes
involved. By contrast, the double-counting to be discussed here requires a
careful re-examination of Taylor's method in order to pinpoint the precise
fault
involved.

\subsection {Two examples of double-counting in the Taylor method}

\subsubsection {Double-counting in pion absorption on the deuteron}

The first example we examine is the double-counting of certain covariant
perturbation theory diagrams in theories of pion absorption on the deuteron, a
problem  first pointed out by Kowalski et al. \cite{Ko79}. Consider two
distinguishable nucleons, which in the initial state are labeled $N1$ and $N2$
and in the final state are labeled $N1'$ and $N2'$, and suppose that also
present in the initial state is a pion, which we label simply $\pi$. We call
the
$3 \rightarrow 2$ amplitude for this process $F$. Consider, in particular,  the
two-particle irreducible part of $F$, $F^{(2)}$. In the discussion here we
ignore the restrictions which should be placed on the amplitudes in the sums of
$C_3$, $C_4$ and $C_5$ (they make no difference to the thrust of the argument).
Consequently Taylor's method and the TRAB simplification thereof are
equivalent.
Both give the following equation:
\begin{equation}
{F}^{(2)}={F}^{(3)} +
\left \{ \left[F^{(3)} + \sum_{j=1,2} f^{(2)}(j)\, d^{-1}_{j}\right]
d_{1} d_{2} d_\pi \left[M^{(2)} + \sum_{i=1,2} t_{\pi N}^{(1)}(i)\,
d^{-1}_{i} + T^{(1)}_{NN}\, d^{-1}_{\pi}\right] \right\}^{(c)},
\label{eq:F2}
\end{equation}
where $M$ is the connected three-to-three amplitude, $t_{\pi N}(i)$ is the
two-body $\pi N$ t-matrix with nucleon $i$ as a spectator, $T_{NN}$ is the
two-body $NN$ t-matrix, $f(j)$ is the $\pi N N$ absorption vertex with particle
$Nj'$ as a spectator and $d_1$, $d_2$ and $d_\pi$ are the single-particle
fully-dressed free propagators for nucleon 1, nucleon 2 and the pion. Note that
the irreducibilities of all amplitudes are indicated by the bracketed
superscript. Note also the use of spectator notation to show the
particles involved in the two-body interaction.  Eq.~(\ref{eq:F2}) is
presented pictorially in Fig.~\ref{Fig.16}, which includes an
indication of the Taylor class or sub-class that produced each term.

Consider, for the moment, only the product of the two-body t-matrices and the
$\pi NN$ vertices. These give a contribution:
\begin {equation}
\sum_{i=1,2}  f^{(2)}(i)\, d_{\overline{i}}\, T^{(1)}_{NN} +
\sum_{i,j=1,2}
\overline{\delta}_{ij} {f}^{(2)}(j)\, d_\pi\, t_{\pi
N}^{(1)}(i)\label{eq:vertt}
\end{equation}
to $F^{(2)}$. Here $\overline {i}$ is defined by:
\begin{equation}
\overline{i}=\left \{ \begin{array}{ll} 2 & \mbox { if $i=1$}\\ 1 &
\mbox { if $i=2$}
\end{array}\ . \right.
\end{equation}

We shall see in the next example that the one-particle irreducible $NN$
t-matrix
contains a term representing one-pion exchange with undressed vertices:
\begin{equation}
f^{(2)*}(2)\, d_\pi\, {f^{(2)*}}^\dagger (1)\ ,
\end{equation}
where $f^{(2)*}$ is two-particle irreducible in all channels. (This term is
depicted in Fig.~\ref{Fig.17}.) It can also be shown that the one-particle
irreducible $\pi N$ t-matrix $t_{\pi N}^{(1)}(i)$ contains a crossed term:
\begin{equation}
f^{(2)}(i)\, d_{\overline{i}}\, {f^{\dagger}}^{(2)} (i)\ .
\end{equation}
(See Fig.~\ref{Fig.18}.) If these portions of the $N N$ and $\pi N$ t-matrices
are substituted into the expression (\ref{eq:vertt}), and the results treated
in
a time-dependent perturbation theory, then both terms in Eq.~(\ref{eq:vertt})
contain the diagram in Fig.~\ref{Fig.19}. That diagram is
double-counted. So, Taylor's method breaks down in this example: at
least one diagram is counted twice.

Note also that in time-ordered perturbation theory there is no double-counting
since the terms involving the $N N$ and $\pi N$ t-matrix contribute to
different
time-orders (compare  Fig.~\ref{Fig.4}, where the vertices are now known to be
two-particle irreducible, with Fig.~\ref{Fig.19}). It is only in a
time-dependent
perturbation theory, such as covariant perturbation theory, that the diagram in
Fig.~\ref{Fig.4} becomes equal to the diagram of Fig.~\ref{Fig.19}, and the
double-counting problem arises.

\subsubsection {Double-counting in the calculation of one-pion exchange}

Another example of double-counting is provided by the calculation of the
one-pion
exchange potential in the $NN \rightarrow NN$ amplitude. Again, suppose we have
two distinguishable nucleons, which in the initial state are labeled $N1$ and
$N2$ and in the final state are labeled $N1'$ and $N2'$. When Taylor's method
is
applied to the amplitude $T^{(2)}$ it gives the following sum of class
$C_4$. (We
do not write the sum of the other classes here as they are irrelevant to the
argument.)
\be
C_4=\fir{2}(1)\, d_\pi\, \tfai{1}(2) + \fir{2}(2)\, d_\pi\, \tfai{1}(1)\ ,
\ee
where, in order to stop diagrams belonging to $C_5$ being included in $C_4$ we
have defined $\tfai{1}(2)$ to be two-particle irreducible in the $N'(\pi N)$
channel. (See Fig.~\ref{Fig.20}.)

However, the vertices $\fir{2}$ are automatically one-particle irreducible
in the
$N(N'\pi)$-channel, since all particles involved are fully dressed. Therefore,
both terms in $C_4$ represent the same diagram, shown in Fig.~\ref{Fig.21}.
Therefore, Taylor's method leads to the double-counting of this diagram in the
calculation of OPE.

\subsection {Why does this double-counting occur?}

These two examples show clearly that double-counting {\em does} arise when
Taylor's method is applied to a time-dependent perturbation theory. The next
question is: {\em why} does it arise?

We begin to answer this question by noting that, in both the cases discussed
above, the double-counting problem occurs in Taylor's class $C_4$. Therefore we
now examine the argument used by Taylor in his attempt to sum $C_4$.

Observe that if the condition:
\begin{equation}
r>m \quad \mbox { or } \quad r \geq n                     \label{eq:lclbd}
\end{equation}
 holds, as it does for each of the two examples above, then Taylor sums
$C_4$ by constructing the sum of the sub-sub-classes
$C_4^{\tilde{s}_f\tilde{t}_i}$ and applying:
\begin{equation} C_4^{s_f t_i}=\sum_{\rm{All} \ {sets} \ \tilde{s}_f \ \rm{\&}
\
\tilde{t}_i} C_4^{\tilde{s}_f
\tilde{t}_i}\ .                                           \label {eq:sum}
\end{equation}
However, what Taylor
should be trying to construct is the {\em union} of all  sub-sub-classes
$C_4^{\tilde{s}_f \tilde{t}_i}$, not the {\em sum}. That is, the correct
formula
is:
\begin{equation}
C_4^{s_f t_i}=\bigcup_{\rm{All} \ {sets} \ \tilde{s}_f \
\rm{\&}\ \tilde{t}_i} C_4^{\tilde{s}_f \tilde{t}_i}\ .
\end{equation}
The operation of summation used in Eq.~(\ref{eq:sum}) is different to that
of set
union, since a diagram which is a member of two different sub-sub-classes
$C_4^{\tilde{s}_f^1 \tilde{t}_i^1}$ and $C_4^{\tilde{s}_f^2
\tilde{t}_i^2}$ is included twice in such a summation, whereas it is
included only once in a set union. Consequently, the identification:
\begin{equation}
\bigcup_{\rm{All} \ {sets} \ \tilde{s}_f \ \rm{\&} \ \tilde{t}_i}
C_4^{\tilde{s}_f \tilde{t}_i}=\sum_{\rm{All} \ {sets} \ \tilde{s}_f
\ \rm{\&} \ \tilde{t}_i}  C_4^{\tilde{s}_f \tilde{t}_i}
\end{equation}
may be made if and only if all the sets $C_4^{\tilde{s}_f \tilde{t}_i}$ are
disjoint. This means that if condition (\ref{eq:lclbd}) holds then Taylor's
method will produce the correct result for $C_4$ if and only if all the
sub-sub-classes $C_4^{\tilde{s}_f \tilde{t}_i}$ are disjoint. If condition
(\ref{eq:lclbd}) holds and the sub-sub-classes are not disjoint then any
diagram which is a member of more than one sub-sub-class will be
double-counted. Similar results hold for $C_3$ and $C_5$: if the
condition (\ref{eq:uc1}) is violated then $C_3$ will not be
summed correctly unless the sub-sub-classes $C_3^{\tilde{t}_i}$ are
disjoint, and if condition (\ref{eq:uc3}) is not satisfied and the
sub-sub-classes $C_5^{\tilde{s}_f}$ are not disjoint then $C_5$ will not be
summed
correctly.

Suppose then  that condition (\ref{eq:lclbd}) holds. What justification is
there
for assuming that these sub-sub-classes are disjoint? Very little, since, as
was
mentioned above in Section \ref{sec-Taylorrev}, one diagram may have many
different latest minimal/maximal $r$-cuts which cut different minimal and
maximal sets of external lines $\tilde{s}_f$ and $\tilde{t}_i$. Such a diagram
will, by the definition of the sub-sub-classes, belong to many of the sets
$C_4^{\tilde{s}_f \tilde{t}_i}$. Therefore, the different
$C_4^{\tilde{s}_f \tilde{t}_i}$ for a fixed $s_f$ and $t_i$ are not necessarily
disjoint and so it is not necessarily true that:
\begin{equation}
C_4^{s_f t_i}=\sum_{\rm{All} \ {sets} \ \tilde{s}_f \ \rm{\&} \
\tilde{t}_i} C_4^{\tilde{s}_f \tilde{t}_i}\ ,
\end{equation}
where the sum is now restricted to those sets $\tilde{s}_f$ and $\tilde{t}_i$
with $s_f$ and $t_i$ members respectively.

However, it is true that the sub-classes $C_4^{s_f t_i}$ (which are defined
to be
the set of all $C_4$ diagrams whose minimal/maximal r-cut intersects $s_f$
final
and $t_i$ initial lines) {\em are} disjoint, and so:
\begin{equation}
C_4=\sum_{s_f t_i} C_4^{s_f t_i}\ ,
\end{equation}
is the right formula for constructing the sum of class $C_4$, once
the correct sums of the sub-classes $C_4^{s_f t_i}$ are known.

The two double-counting problems above provide perfect examples of this
difficulty. Consider the first example. For the diagram which is double-counted
in this example we have $s_f=t_i=1$ (see Fig.~\ref{Fig.22}). But, there are
four
possible pairs of sets $\tilde{s}_f$ and $\tilde{t}_i$. Two of these four pairs
are:
\begin{eqnarray}
\tilde{s}_f=\{N1'\}\ &;&\quad\tilde{t}_i=\{N2\}\ ,\\
\tilde{s}_f=\{N2'\}\ &;&\quad\tilde{t}_i=\{\pi\}\ .
\end{eqnarray}
For each {\em pair of sets} we may construct a unique latest three-cut, as
shown
in Fig.~\ref{Fig.22}. However, when we attempt to construct the {\em overall
unique latest three-cut} via the technique used in the last internal cut lemma,
we fail because the constructed ``latest" cut, $c^+$, does not constitute
a cut at all, since neither of the lines $N1'$ or $N2'$ is an internal line. As
explained above, it was in order to circumvent this difficulty in the
construction of a unique latest $r$-cut that Taylor constructed the sum of the
individual sub-sub-classes $C_4^{\tilde{s}_f \tilde{t}_i}$ and then summed
over all
possible sub-sub-classes. However, diagrams such as Fig.~\ref{Fig.22} belong to
more than one sub-sub-class (in this case $C_4^{\{N1'\}\{N2\}}$ and
$C_4^{\{N2'\}\{\pi\}}$) and so are counted twice in such a summation over
sub-sub-classes. Consequently, in this case, Taylor's method does not
accurately sum
class $C_4$.

Similarly in the second example a unique latest three-cut cannot be
found (see Fig.\ref{Fig.23}). However, the double-counted diagram belongs to
both
$C_4^{\{N1'\}\{N2\}}$ and $C_4^{\{N2'\}\{N1\}}$ and so is double-counted
when the
sum of $C_4$ is constructed by the methods advocated by Taylor. Again,
Fig.~\ref{Fig.23} shows the impossibility of constructing a unique last
three-cut in this situation, but the problem is {\em not}, as is claimed by
Taylor, solved by summing over the sub-sub-classes
$C_4^{\tilde{s}_f \tilde{t}_i}$, since that summation merely leads to the
double-counting of this diagram, and, indeed, of any other diagram
which belongs to more than one sub-sub-class.

\section {Solving the double-counting problem in the Taylor method}
\label {sec-Dcsoln}

Having discovered this problem with the Taylor method we now attempt
to solve it. In this section we construct a systematic  solution to the type of
double-counting discussed in the previous section.

Firstly note that, as discussed above, this type of double-counting does not
occur in summing over the sub-classes $C_3^{t_i}$, $C_4^{s_f t_i}$ and
$C_5^{s_f}$. Once the correct sums of these sub-classes are obtained the
formulae (\ref{eq:C3sum1}) or (\ref{eq:C3sum2}), (\ref{eq:C4sum}) and
(\ref{eq:C5sum1}) or (\ref{eq:C5sum}) actually give the right result for the
sums of $C_3$,
$C_4$ and $C_5$, since these sub-classes are, by definition, disjoint. The
problem occurs in obtaining the sums of these sub-classes in the first place,
since the sub-sub-classes of these sub-classes: $C_3^{\tilde{t}_i}$,
$C_4^{\tilde{s}_f
\tilde{t}_i}$ and $C_5^{\tilde{s}_f}$ are {\em not} disjoint. In contrast to
the
double-counting fixed in Section~\ref{sec-Taylorrev} this is {\em
intra-}sub-class double-counting.

Now, suppose that $C^l$ is a sub-class of $C_3$, $C_4$ or $C_5$, in which
we have
a certain minimum and/or maximum number of lines from the initial/final state
cut, and that:
\be
C^l=\bigcup_{j=1}^N C^{\tilde{l}_j}\ ,
\ee
where $\{\tilde{l}_j:j=1,2,\ldots,N\}$ is a set of $N$ sets of external lines,
all with $l$ members. Then it is clear that the sum of $C^l$ may be found by
taking each of the sub-sub-classes in turn and adding them one at a time to a
``running sum". However, a particular sub-sub-class $C^{\tilde{l}_j}$ may only
be added to this running sum after any diagrams already included in the running
sum have been removed from it. Mathematically this procedure is expressed as
follows:
\be
C^l=\sum_{j=1}^N  C^{\tilde{l}_j} - \sum_{k<j}
 C^{\tilde{l}_j} \cap C^{\tilde{l}_k} + \sum_{h<k<j} C^{\tilde{l}_j} \cap
C^{\tilde{l}_k} \cap C^{\tilde{l}_h}  - \ldots \ .
\label {eq:Clwithcorrn}
\ee
Taylor merely ignores all but the first term in this expression, and that is
what leads to the intra-sub-class double-counting discussed in the previous
section.  Note once again that the difference between
this type of double-counting and that discussed in
Section~\ref{sec-Taylorrev} is
that above we dealt with double-counting between different classes, or
between the different sub-classes $C^{l}$ (inter-class or inter-sub-class
double-counting). Here the double-counting is between the sub-sub-classes
$C^{\tilde{l}_j}$ of a particular sub-class $C^l$ (intra-sub-class
double-counting).

Before dealing specifically with $C_3$--$C_5$ we observe that
diagrams in the intersection $C^{\tilde{l}_j} \cap C^{\tilde{l}_k}$
will be those which admit two different $r$-cuts: one involving the
set of external lines $\tilde{l}_j$, and one involving the set of
external lines $\tilde{l}_k$. Consequently by examining the result
for $C^{\tilde{l}_j}$ and determining which diagrams in that sum
admit an $r$-cut involving the set of lines $\tilde{l}_k$ we may
determine the sum of the diagrams in $C^{\tilde{l}_j} \cap
C^{\tilde{l}_k}$. This procedure involves examining the overall
$s$-channel cut-structure of the sum of $C^{\tilde{l}_j}$. In
time-ordered perturbation theory this cut-structure is fully determined
by the $s$-channel cut-structure of the sub-amplitudes making up the full
amplitude. However, in any time-dependent perturbation theory the
$s$-channel cut-structure depends, not only on the $s$-channel cut
structure of the sub-amplitudes, but also on their cut-structure in
other channels. Therefore, in order to eliminate double-counting it is
necessary to examine the cut-structure of the sub-amplitudes in a
number of different channels.

Similar considerations allow us to (if necessary) construct the
intersections of
three or more sub-classes. That is to say, the procedure for evaluating such
intersections is merely an extension of that for finding $C^{\tilde{l}_j} \cap
C^{\tilde{l}_k}$. Consequently, we now apply the ideas of the previous
paragraph to each class $C_3$--$C_5$ in turn, thus showing how to calculate the
first correction term in Eq.~(\ref{eq:Clwithcorrn}) for any sub-class of
$C_3$--$C_5$. We leave it to the interested reader to extend this argument
to the
calculation of further correction terms in Eq.~(\ref{eq:Clwithcorrn}).

\subsection {$C_3$}

In this case we wish to calculate:
\be
\ctt{j}\cap \ctt{k}\ ,
\ee
where $\tit{j}$ and $\tit{k}$ are two minimal sets of initial-state
lines, in order to obtain the corrected sum of $C_3^{t_i}$.

As was explained above, if $2t_i < m$ then even if $r>m$ the sum of any diagram
in $C_3^{t_i}$ may still be constructed by finding the unique latest cut in
$M_{t_i}$. Thus unless $2t_i \geq m$ and $r>m$ it is not necessary to pursue
the
construction of sub-sub-classes $C_3^{\tilde{t}_i}$. Consequently the
following discussion only applies to the case $2t_i \geq m$. If $2t_i\geq m$ it
follows that:
\be
\tit{j} \cap \tit{k} \neq \phi\ .
\ee

Now, any diagram in $\ctt{j}\cap \ctt{k}$ admits two $r$-cuts $X$ and
$Y$:
\bea
X&=&\tit{j} \cup \{r-t_i \mbox{ internal lines}\}\\
Y&=&\tit{k} \cup \{r-t_i \mbox{ internal lines}\}\ ,
\eea
which are, respectively, the latest $r$-cuts involving the sets of lines
$\tit{j}$ and $\tit{k}$. The set of initial-state lines $(m)$ may be
written as:
\be
(m)=(m/\tit{j}) \cup (m/\tit{k}) \cup (\tit{j} \cap \tit{k})\ ,
\ee
therefore if we define:
\be
\tilde{I}=(m/\tit{j}) \cap (m/\tit{k})
\ee
we have:
\be
(m)=[(m/\tit{j})/\tilde{I}] \cup [(m/\tit{k})/\tilde{I}] \cup
\tilde{I} \cup (\tit{j} \cap \tit{k})\ ,
\label{eq:mdecomp}
\ee
where all four sets in this union are disjoint.
But:
\be
(m)=(m/\tit{j}) \cup \tit{j}\ ,
\ee
and comparing this equation with Eq.~(\ref{eq:mdecomp}) gives:
\be
\tilde{t}_i^{(j)}=[(m/\tit{k})/\tilde{I}] \cup
(\tit{j} \cap \tit{k})\ .
\label{eq:titj}
\ee
(Similarly for $\tit{k}$.) Note that if we write $N(\tilde{I})=I$ we
have: \bea
N(\tit{j} \cap \tit{k})=2t_i-m+I\\
N([(m/\tit{j})/\tilde{I}])=m-t_i-I\ .
\eea
Eq.~(\ref{eq:titj}) then suggests that:
\bea
X&=&[(m/\tit{k})/\tilde{I}] \cup (\tit{j} \cap \tit{k}) \cup \{r-t_i
\mbox{ internal lines}\},\\
Y&=&[(m/\tit{j})/\tilde{I}] \cup (\tit{j} \cap \tit{k}) \cup \{r-t_i
\mbox{ internal lines}\}\ .
\eea
Eliminating the set $\tit{j} \cap \tit{k}$ from consideration
produces two $(r-2t_i+m-I)$-cuts, $X'$ and $Y'$, which are to be made
on a $(2m-2t_i-I) \rightarrow n$ diagram.

The LICL procedure is now used to construct $c^+_{X' Y'}$ and
$c^-_{X' Y'}$. The argument given in Section \ref{sec-Class3} above suggests
that $c^+_{X' Y'}$ will be an $(r-2t_i+m-I)$-cut, provided $c^-_{X' Y'}$ does
not contain only initial-state lines. But, the only initial-state lines which
could be in $c^-_{X' Y'}$ are those in the set:
\be
S=[(m/\tit{j})/\tilde{I}] \cup [(m/\tit{k})/\tilde{I}]\ .
\ee
Since these two sets are disjoint:
\be
N(S)=2m-2t_i-2I < 2m-2t_i-I,
\ee
provided $\tilde{I} \neq \phi$. Consequently, if $\tilde{I} \neq \phi$,
$c^{-}_{X'Y'}$ cannot consist solely of initial-state lines, and so
constructing:
\be
c^+_{XY}=c^+_{X'Y'} \cup (\tit{j} \cap \tit{k})
\ee
yields an $r$-cut which is later than both $X$ and $Y$. Such an
$r$-cut may only cut initial-state lines cut by both $X$ and
$Y$. It follows that $c^+_{XY}$ cuts only initial-state lines from
$\tit{j} \cap \tit{k}$, i.e. it cuts only $(2t_i - m+I)$ initial-state
lines. But:
\be
2t_i-m+I=t_i-(m-t_i-I) < t_i\ ,
\ee
therefore, in this case any diagrams which may be in $\ctt{j} \cap
\ctt{k}$ actually belong to $C_3^{2t_i-m+I}$, not to $C_3^{t_i}$,
which is a contradiction. Thus, it follows that we must have:
\be
\tilde{I}=\phi\ ,
\ee
if $\ctt{j} \cap \ctt{k} \neq 0$. Adapting Eq.~(\ref{eq:titj})
therefore shows that if $\ctt{j} \cap \ctt{k} \neq 0$ we must have:
\bea
\tit{j}=(m/\tit{k}) \cup (\tit{j} \cap \tit{k})\ ;\\
\tit{k}=(m/\tit{j}) \cup (\tit{j} \cap \tit{k})\ .
\eea

Once these two equations are derived we are in a position to examine
the diagrammatic result for the sum of $\ctt{j}$ and derive, using
the LICL, the result for $\ctt{j} \cap \ctt{k}$ shown in
Fig.~\ref{Fig.24}. We write this result algebraically as:
\be
\ctt{j} \cap \ctt{k}=\left[\,{\A{(2r-m)}{n}{r}}_{\tit{j} \tit{k}}
\ G^{(r/\tit{j})}\ \A{(m-t_i)}{(r-t_i)}{r-t_i-1}\ G^{(r/\tit{k})}
\ \A{(m-t_i)}{(r-t_i)}{r-t_i-1}\,\right]^{(c)}\ ,      \label{eq:C3int}
\ee
if $(m/\tit{j}) \cap (m/\tit{k}) = \phi$ and:
\be
\ctt{j} \cap \ctt{k}=0
\label{eq:C3int2}
\ee
otherwise.

Note the following points about this result:
\begin{enumerate}
\item The amplitude $\A{(2r-m)}{n}{r}$ is a more complicated amplitude
than $\A{r}{n}{r}$, since $r >m$ is a necessary condition for
double-counting.

\item Furthermore, the notation ${\A{(2r-m)}{n}{r}}_{\tit{j} \tit{k}}$
indicates that $\A{(2r-m)}{n}{r}$ has had the following constraints
placed on it:
\begin{enumerate}
\item Constraints to stop $r$-cuts involving any number of final-state
lines, or cuts intersecting less than $r$ lines. In fact, the
constraint of
$(r-1-m+t_i)$-particle irreducibility imposed in the channels:
\[
[n/\tilde{h}] + [r/\tilde{t}_i^{(k)}] \leftarrow
[\tilde{h}] + [\tit{j} \cap \tit{k}] + [r/\tit{j}]
\]
and
\[
[n/\tilde{h}] + [r/\tilde{t}^{(j)}_i] \leftarrow
[\tilde{h}] + [\tit{j} \cap \tit{k}] + [r/\tilde{t}^{(k)}_i]\ ,
\]
where $\tilde{h}$ is any one final-state line, is enough to preclude
the possibility of such cuts.

\item $(r-t_i)$-particle irreducibility in the channels:
\[
[\tit{j} \cap \tit{k}] +  [r/\tilde{t}^{(k)}_i] + [n]
\leftarrow [r/\tilde{t}^{(j)}_i]
\]
and
\[
[\tit{j} \cap \tit{k}] +  [r/\tilde{t}^{(j)}_i] + [n]
\leftarrow [r/\tilde{t}^{(k)}_i] \ .
\]
\end{enumerate}

In fact, both these constraints may be enforced by requiring that
$\A{(2r-m)}{n}{r}$ be $(r-m+t_i)$-particle irreducible in  the
channels:
\[
[n] + [r/\tilde{t}^{(k)}_i] \leftarrow
[\tit{j} \cap \tit{k}] + [r/\tit{j}]
\]
and
\[
[n] + [r/\tilde{t}^{(j)}_i] \leftarrow
[\tit{j} \cap \tit{k}] + [r/\tit{k}]
\]
as indicated in Fig.~\ref{Fig.24}.
\end{enumerate}

Given this result for $\ctt{j} \cap \ctt{j}$, if $r>m$ and $2t_i
\geq m$ the first correction to the sum of
$C_3^{t_i}$ may be calculated
via the formula:
\be
C_3^{t_i}=\sum_{j=1}^p
\left(\ctt{j}-\sum_{k=1}^{j-1}
\ctt{j} \cap \ctt{k} + \ldots\right),
\ee
where $p={\small {\comb{m}{t_i}}}$, the sum in the first term is written in
Eq.~(\ref{eq:C3ti2}) and
$\ctt{j} \cap \ctt{k}$ is given by Eqs.~(\ref{eq:C3int}) and
(\ref{eq:C3int2}).

\subsection{$C_4$}

We begin by observing that if $2t_i<m$ and $2s_f<n$ the formula
(\ref{eq:C42}) gives the correct result for $C_4^{s_f t_i}$.
Furthermore, from our discussion in Section \ref{sec-Class4} above we know that
we cannot have $(2t_i-m+n-2s_f) > 0$. However, if $2t_i \geq m$ or $2s_f \geq
n$
while $(2t_i-m+n-2s_f) \leq 0$ we must use the formula:
\be
C_4^{s_f t_i}=\sum_{j=1}^q\
\left[\,\cfost{j} - \sum_{k=1}^{j-1} \cfost{j} \cap \cfost{k} +
\ldots \,\right]\ .
\ee
where $q={\small{\comb{n}{s_f} \times \comb{m}{t_i}}}$.
We now calculate the first correction term here for the three different cases
which can occur given these conditions:

\subsubsection*{Case 1: $2t_i \geq m$ and $2s_f<n$}

Once again we define:
\be
(m/\tit{j}) \cap (m/\tit{k})=\tilde{I}\ .
\ee
We originally assume that $\tilde{I} \neq \phi$ and construct the two
$r$-cuts:
\bea
X&=&\tit{j} \cup \{r-s_f-t_i \mbox{ internal lines}\} \cup \sft{j}
\label{eq:A}\\
Y&=&\tit{k} \cup \{r-s_f-t_i \mbox{ internal lines}\} \cup
\sft{k}\ ,                                         \label{eq:B}
\eea
which are, respectively, the two latest $r$-cuts cutting the sets of
external lines $\tit{j} \cup \sft{j}$ and $\tit{k} \cup \sft{k}$.
Using a similar argument to that applied above we may then show that
an $r$-cut $c^+_{XY}$ may be constructed which is later than both $X$
and $Y$. As above, this leads to a contradiction. Note that in this
case it must be checked that $c^+_{XY}$ does not consist entirely of
final-state lines. This, however, is guaranteed by the condition
$2s_f < n$. Thus,
\be
\tilde{I}=\phi                                 \label{eq:Icond}
\ee
is a necessary condition if $\cfost{j} \cap \cfost{k} \neq 0$.

We must also consider whether or not $\sft{j} \cap \sft{k}=\phi$.
Examination of the diagram representing the sum of
$\cfost{j}$ shows that the existence of an $r$-cut involving
$\sft{k}$ and $\tit{k}$ with $\sft{j} \cap \sft{k} \neq \phi$ must also
imply that a cut cutting less than $r$ lines may be made on the
diagram (see Fig.~\ref{Fig.25}). Therefore, if $\cfost{j}
\cap \cfost{k} \neq 0$ we must have:
\be
\sft{j} \cap \sft{k} = \phi\ .
\label{eq:sfint}
\ee

Once Eqs.~(\ref{eq:Icond}) and (\ref{eq:sfint}) are established
applying the LICL to the sum of $\cfost{j}$ implies that:
\bea
\cfost{j} \cap \cfost{k}=\Bigg[\,{\A{(2r-m-2s_f)}{(n-2s_f)}{r-2s_f}}_{\tit{j}
\tit{k}}\ G^{(r/(\tit{j} \cup \sft{j}))}\
{\A{(m-t_i)}{(r-t_i)}{r-t_i-1}}_{\tilde{s}_f^{(j)}}
\nn\\
\times\ G^{(r/(\tit{k}\cup \sft{k}))}\
{\A{(m-t_i)}{(r-t_i)}{r-t_i-1}}_{\tilde{s}_f^{(k)}}\,\Bigg]^{(c)}\ ,
\eea
if $(m/\tit{j}) \cap (m/\tit{k})=\phi$ and $\sft{j} \cap
\sft{k}=\phi$, while:
\be
\cfost{j} \cap \cfost{k}=0\ ,
\ee
otherwise. This result is depicted in Fig.~\ref{Fig.26}.

Note the following facts about this result:
\begin{enumerate}
\item ${\A{(m-t_i)}{(r-t_i)}{r-t_i-1}}_{\tilde{s}_f}$ has the same
restrictions on it as in Section~\ref{sec-Class4}.

\item ${\A{(2r-m-2s_f)}{(n-2s_f)}{r-2s_f}}_{\tit{j}
\tit{k}}$ must be $(r-m+t_i-s_f)$-particle irreducible in the two
$[r-t_i-s_f] + [n-2s_f] \leftarrow [r-t_i-s_f] + [2t_i - m]$-channels indicated
in Fig.~\ref{Fig.26}.
\end{enumerate}

\subsubsection*{Case 2: $2t_i<m$ and $2s_f \geq n$}

Essentially, in this case the roles of initial and final-state lines
in the above argument are reversed. We construct:
\be
(n/\sft{j}) \cap (n/\sft{k})=\tilde{I}\ ,
\ee
and show:
\be
\sft{k}=[(n/\sft{j})/\tilde{I}] \cup (\sft{j} \cap \sft{k})\ ,
\ee
similarly for $\sft{j}$. Then using the same type of argument
employed above we show that the two $r$-cuts $X$ and $Y$ defined by
Eqs.~(\ref{eq:A}) and (\ref{eq:B})  can only be possible if:
\bea
\tilde{I}&=&\phi\\
\tit{j} \cap \tit{k}&=&\phi\ .
\eea
It follows that if these two conditions are not satisfied then
$\cfost{j} \cap \cfost{k}=0$. If they are satisfied then the LICL,
applied to the sum of $\cfost{j}$, implies that:
\bea
\cfost{j} \cap
\cfost{k}\Bigg[\,{\A{(r-s_f)}{(n-s_f)}{r-s_f}}_{\tilde{t}_i^{(k)}}\
G^{(r/(\tit{k} \cup
\sft{k}))}\ {\A{(r-s_f)}{(n-s_f)}{r-s_f}}_{\tilde{t}_i^{(j)}}
\ G^{(r/(\tit{j} \cup \sft{j}))}\nn\\
\times\ {\A{(m-2t_i)}{(2r-n-2t_i)}{r-2t_i-1}}_{\sft{j} \sft{k}}\,\Bigg]^{(c)}\
,
\eea
a result shown diagrammatically in Fig.~\ref{Fig.27}.
Here:
\begin{enumerate}
\item ${\A{(r-s_f)}{(n-s_f)}{r-s_f}}_{\tilde{t}_i}$ has the
constraint discussed in Section~\ref{sec-Class4} for the case $2t_i < m$
imposed on it.

\item ${\A{m-2t_i}{2r-n-2t_i}{r-2t_i-1}}_{\sft{j} \sft{k}}$ must be
$(r-s_f-t_i-1)$-particle irreducible in the two channels:
\[
[r-t_i-s_f] \leftarrow [r-t_i-s_f] + [m-2t_i] + [2s_f-n]
\]
as is indicated in Fig.~\ref{Fig.27}.
\end{enumerate}

\subsubsection*{Case 3: $2s_f=n$ and $2t_i=m$}

Now consider the case $2s_f=n$ and $2t_i=m$. Once again we may show
that if the two $r$-cuts $X$ and $Y$, defined by (\ref{eq:A}) and
(\ref{eq:B}) are both to be possible on a diagram in $C_4^{s_f t_i}$ then we
must have:
\bea
\sft{j} &\cap& \sft{k}=\phi\ , \label{eq:int1}\\
\tit{j} &\cap& \tit{k}=\phi\ ,  \label{eq:int2}
\eea
which, in this case, suggests that:
\bea
(n/\sft{j})&=&\sft{k}\ ,\\
(m/\tit{j})&=&\tit{k}\ .
\eea
Once this is known it is clear that the $r$-cut $Y$ is the cut
indicated in Fig.~\ref{Fig.28}, which represents the sum of
$\cfost{j}$, complete with constraints to stop $r$-cuts involving final-state
lines and cuts cutting less than $r$-lines. Therefore, the whole of $\cfost{j}$
is double-counted, i.e.~if the conditions~(\ref{eq:int1}) and (\ref{eq:int2})
are obeyed we have:
\be
\cfost{j} \cap \cfost{k}=\cfost{j}=\cfost{k}\ .
\ee

If the figure representing $\cfost{j}$ or $\cfost{k}$ is redrawn
it becomes clear that the
two diagrams for $\cfost{j}$ and $\cfost{k}$ are, in fact, the same,
and so this entire diagram is, indeed, included in both sub-sub-classes.
(See Fig.~\ref{Fig.29}.)

\subsection{$C_5$}

In order to calculate the correct sum of $C_5$ it is necessary to use
the equation:
\be
C_5^{s_f}=\sum_{j=1}^\ell \ \left[\,\cfs{j}-\sum_{k=1}^j
\cfs{j}\cap \cfs{k} + \ldots \,\right]\ ,
\ee
with $\ell={\small{\comb{n}{s_f}}}$,
whenever $2s_f \geq n$ and $r \geq n$. If $2s_f < n$ or $r < n$ then, as was
described above, the whole process of breaking
$C_5^{s_f}$ into sub-sub-classes is unnecessary.

The argument for the construction of $\cfs{j} \cap \cfs{k}$ is again
similar to that used previously. Since $2s_f \geq n$ we have:
\be
\sft{j} \cap \sft{k} \neq \phi\ ,
\ee
and once more we construct:
\be
(n/\sft{j}) \cap  (n/\sft{k}) \equiv \tilde{I}\ ,
\ee
and show that unless $\tilde{I}=\phi$ the two ``latest" maximal
$r$-cuts: \bea
X&=&\{r-s_f \mbox{ internal lines}\} \cup \sft{j}\\
Y&=&\{r-s_f \mbox{ internal lines}\} \cup \sft{k}\ ,
\eea
cannot both be made on the diagram, as if $\tilde{I} \neq \phi$ and
both $X$ and $Y$ are possible then an $r$-cut involving more than
$s_f$ final-state lines is also possible. Therefore,
\be
(n/\sft{j}) \cap (n/\sft{k}) \neq \phi
\implies \cfs{j} \cap \cfs{k}=0
\ee
but if $(n/\sft{j}) \cap  (n/\sft{k}) = \phi$ then the LICL
may be used to show that:
\begin{eqnarray}
\cfs{j} \cap \cfs{k}&=&\Bigg[\,\A{(r-s_f)}{(n-s_f)}{r-s_f}
\ G^{(r/(\tit{k} \cup \sft{k}))}\
\A{(r-s_f)}{(n-s_f)}{r-s_f}\nonumber\\
& &\qquad\qquad\qquad\times\  G^{(r/(\tit{j} \cup
\sft{j}))}\ {\A{m}{(2r-n)}{r-1}}_{\sft{j} \sft{k}}\,\Bigg]^{(c)}\ ,
\end{eqnarray}
where ${\A{m}{(2r-n)}{r-1}}_{\sft{j} \sft{k}}$ has the restriction of
$(r-s_f-1)$-particle irreducibility in the two channels:
\[
[r-s_f]  \leftarrow [m] + [r-s_f] + [2s_f-n]\ .
\]
(See Fig.~\ref{Fig.30}.)

\section {Conclusion}

In this paper we have reexamined the Taylor method of
classification-of-diagrams. A review of the classification-of-diagrams
scheme has been given and two questions regarding this method have
been answered.

The first question involved the simplification of the Taylor method
developed by Thomas, Rinat, Afnan and Blankleider~\cite{TR79,AB80}. We
found that this method, which was originally derived for use in
time-ordered perturbation theory, is, in fact, equivalent to the
Taylor method in time-dependent perturbation theory too, provided that
when using the full Taylor method we ignore the constraints which
should be imposed on the cut-structure of sub-amplitudes in channels
other than the $s$-channel. This explains how Afnan and Blankleider,
who used the TRAB method in their work on the $NN-\pi NN$
system~\cite{AB80} still managed to obtain the equations found by
Avishai and Mizutani using the ``full" Taylor method~\cite{AM83}.

Secondly, we showed that the Taylor method double-counts certain
diagrams when it is applied in a time-dependent perturbation theory.
We found that this double-counting can occur in two ways:
\begin {enumerate}
\item {\bf Inter-class or inter-sub-class double-counting:} While reviewing the
Taylor method we discovered that certain diagrams are included in more than one
class, $C$, or sub-class, $C^n$, unless constraints are placed on the
amplitudes
in channels other than  the $s$-channel. However, if these constraints are
imposed this type of double-counting is eliminated. Note that certain diagrams
which are not $(r-1)$-particle irreducible may also be (incorrectly) included
in
the sum of $\A{m}{n}{r-1}$ unless such restrictions are applied.

\item {\bf Intra-sub-class double-counting:} Taylor's division of all diagrams
within a sub-class
$C^n$ into sub-sub-classes $C^{\tilde{n}}$ places some diagrams in more
than  one
sub-sub-class. These diagrams are then double-counted
when the sums of all the different sub-sub-classes are added together. We
outlined a general procedure by which this type of double-counting
 can be eliminated. We then used Taylor's
own LICL to calculate expressions for part of the sum of the double-counted
diagrams.
\end {enumerate}

The double-counting-removal techniques developed in this
paper are equivalent to examining the full $s$-channel cut-structure
of the amplitude in question and placing constraints on the cut-structure of
the
sub-amplitudes contributing to this amplitude in channels other than the
$s$-channel in order to eliminate the double-counting.  By contrast, Taylor's
original method does not sufficiently constrain the cut-structure of the
sub-amplitudes in these other channels---it (almost exclusively) only
constrains their
$s$-channel cut-structure. From a topological point of view this
under-specification of the cut-structure in channels other than the
$s$-channel is the reason why Taylor's method leads to double-counting
when it is applied in a time-dependent perturbation theory.

The modified Taylor method developed in this paper may now be used in
order to derive double-counting-free integral equations for systems of
mesons and baryons. In particular, these ideas will be applied to the
derivation of equations for the amplitudes in a covariant theory of
nucleons and pions in a forthcoming paper~\cite{AP93C}. This results
in equations for the $NN-\pi NN$ system which are covariant and
free from the double-counting problems of previous four-dimensional
equations.

\acknowledgements {We would both like to acknowledge A.~N.~Kvinikhidze
and B.~Blankleider, each of whom gave us a hint that something might
be wrong with the Taylor method. D.~R.~P. wishes to acknowledge
discussions with S.~B.~Carr. We are also very grateful to our
referee, whose perceptive comments contributed significantly to the
content of this paper. We both wish to thank the Australian Research
Council for their financial support. D.~R.~P. holds an Australian
Postgraduate Research Award.}

\tighten
\begin {figure}
\vspace{35 mm}
\hskip 5 mm
\special {illustration :Diagrams:Fig.1}
\vskip 2 mm
\caption {Two diagrams, both of which give contributions to pion
absorption on the deuteron in the Taylor method.}
\label{Fig.1}
\end {figure}

\begin {figure}
\vspace{18 mm}
\hskip 42 mm
\special {illustration :Diagrams:Fig.2}
\vskip 2 mm
\caption {The crossed term in the $\pi N$ t-matrix.}
\label {Fig.2}
\end {figure}

\begin {figure}
\vspace{30 mm}
\hskip 40 mm
\special {illustration :Diagrams:Fig.3}
\vskip 2 mm
\caption {The diagram which Kowalski et al.~\protect{\cite{Ko79}} pointed
out was
double-counted in certain models of pion absorption on the deuteron.}
\label{Fig.3}
\end {figure}

\begin {figure}
\vspace{30 mm}
\hskip 40 mm
\special {illustration :Diagrams:Fig.4}
\vskip 2 mm
\caption {The diagram obtained when the crossed term is substituted
into the right-hand diagram of Fig.~\protect{\ref{Fig.1}} in time-ordered
perturbation theory. Note that, in time-ordered perturbation theory, this
diagram
is {\em not} included in the left-hand diagram of Fig.~\protect{\ref{Fig.1}}.}
\label{Fig.4}
\end {figure}

\vfill
\eject
\null

\begin{figure}
\vskip 15 cm
\special{illustration :Diagrams:Fig.5}
\caption{The classification of some $(r-1)$-particle irreducible $n \leftarrow
m$ diagram into one of the classes $C_1$ -- $C_5$ according to the
kinds of $r$-cut which may be made on it.}\label{Fig.5}
\end{figure}

\vfill
\newpage
\null

\begin {figure}
\vspace{40 mm}
\hskip 5 mm
\special {illustration :Diagrams:Fig.6}
\vskip 3 mm
\caption {Two of the ways in which two cuts, $c_1$ and $c_2$, may
intersect, and the resulting definitions of $c^+=c_1^+ \cup c_2^+$ and
$c^-=c_1^- \cup c_2^-$ in each of the two cases.}
\label{Fig.6}
\end {figure}

\begin {figure}
\vspace{50 mm}
\hskip 8 mm
\special {illustration :Diagrams:Fig.7}
\vskip 3 mm
\caption {A diagrammatic representation of the sum of Taylor class
$C_2$.}
\label {Fig.7}
\end {figure}

\begin {figure}
\vspace{55 mm}
\hskip 38 mm
\special {illustration :Diagrams:Fig.8}
\caption {An example for the case $n=3$, $m=r=4$ in which parts of the
expression derived for $C_2$ are not $(r-1)$-particle irreducible if the
disconnected pieces of amplitudes are not carefully defined.}
\label {Fig.8}
\end {figure}

\vfill
\eject
\null

\begin {figure}
\vspace{45 mm}
\hskip 15 mm
\special {illustration :Diagrams:Fig.9 scale 900}
\vskip 3 mm
\caption{Two cuts $\alpha$ and $\beta$ which, if they are $r$-cuts,
will place this diagram, summed in $C_3^{t_i}$, in, respectively,
$C_4$, or (if $2t_i > m$) a different sub-class of $C_3$. Note that $\alpha$
and $\beta$ could also cut less than $r$ lines.}
\label {Fig.9}
\end {figure}

\begin {figure}
\vspace{45 mm}
\hskip -5 mm
\special {illustration :Diagrams:Fig.10 scaled 800}
\caption {A diagrammatic representation of the sum of Taylor class
$C_3$.}
\label {Fig.10}
\end {figure}

\begin {figure}
\vspace{60 mm}
\hskip 15 mm
\special {illustration :Diagrams:Fig.11}
\vskip 1 mm
\caption {A diagram in $C_4^{s_f t_i}$, with a cut which cuts less
than $r$ lines if $(2t_i-m+2s_f-n) > 0$.}
\label {Fig.11}
\end {figure}

\vfill
\eject
\null

\begin {figure}
\vspace{65 mm}
\hskip 20 mm
\special {illustration :Diagrams:Fig.12}
\vskip 3 mm
\caption {Two cuts $\alpha$ and $\beta$ which, if they are $r$-cuts,
will place this diagram, summed in $C_4^{s_f t_i}$, in, respectively,
$C_5$, or (if $2t_i > m$) a different sub-class of $C_4$. Note also that these
two cuts could cut less than $r$ lines.}
\label {Fig.12}
\end {figure}

\begin {figure}
\vspace{70 mm}
\hskip -23 mm
\special {illustration :Diagrams:Fig.13}
\vskip 3 mm
\caption {A diagrammatic representation of the sum of Taylor class
$C_4$.}
\label {Fig.13}
\end {figure}

\vfill
\eject
\null

\begin {figure}
\vspace{60 mm}
\hskip 20 mm
\special {illustration :Diagrams:Fig.14}
\vskip 3 mm
\caption {A cut which, if it is an $r$-cut, will place this diagram,
summed in $C_5^{s_f}$, in a different sub-class of $C_5$, if $2s_f
< n$. Note also that this cut could intersect fewer than $r$ lines.}
\label {Fig.14}
\end {figure}

\begin {figure}
\vspace{65 mm}
\hskip -17 mm
\special {illustration :Diagrams:Fig.15}
\caption {A diagrammatic representation of the sum of Taylor class
$C_5$.}
\label {Fig.15}
\end {figure}

\vfill
\eject
\null

\begin {figure}
\vspace{200 mm}
\hskip -5 mm
\special {illustration :Diagrams:Fig.16}
\vskip 3 mm
\caption {The equation for the two-particle irreducible $\pi NN$ to
$NN$ amplitude, $F^{(2)}$, which is obtained from Taylor's method,
with the Taylor classes or sub-sub-classes which produce each term
indicated.}
\label {Fig.16}
\end {figure}

\vfill
\eject
\null

\begin {figure}
\vspace{20 mm}
\hskip 50 mm
\special {illustration :Diagrams:Fig.17}
\vskip 3 mm
\caption {Part of the $NN$ t-matrix.}
\label {Fig.17}
\end {figure}

\begin {figure}
\vspace{20 mm}
\hskip 46 mm
\special {illustration :Diagrams:Fig.18}
\vskip 3 mm
\caption {The crossed term in the $\pi N$ t-matrix, with the
irreducibility of each vertex indicated.}
\label {Fig.18}
\end {figure}

\begin {figure}
\vspace{35 mm}
\hskip 45 mm
\special {illustration :Diagrams:Fig.19}
\vskip 3 mm
\caption {One diagram which is double-counted when the Taylor method
is used to derive an equation for $F^{(2)}$.}
\label {Fig.19}
\end {figure}

\begin {figure}
\vspace{38 mm}
\hskip -5 mm
\special {illustration :Diagrams:Fig.20}
\vskip 3 mm
\caption {The two diagrams contributing to $C_4$ for $T^{(2)}$ in the
Taylor method, with the two cuts which will place diagrams in
$C_5$ if they are three-cuts.}
\label{Fig.20}
\end {figure}

\vfill
\eject
\null

\begin {figure}
\vspace{38 mm}
\hskip 47 mm
\special {illustration :Diagrams:Fig.21 scaled 1200}
\vskip 3 mm
\caption {The diagram which both $C_4^{\{N1'\} \{N2\}}$ and
$C_4^{\{N2'\} \{N1\}}$ sum to when the Taylor method is applied to
$T^{(2)}$.}
\label{Fig.21}
\end {figure}

\begin {figure}
\vspace{50 mm}
\hskip 22 mm
\special {illustration :Diagrams:Fig.22}
\vskip 3 mm
\caption {The two possible ``latest" cuts, $c_1$ and $c_2$, which lead
to the double-counting of Fig.~
\protect{\ref{Fig.19}}, and the ``cuts", $c^-$ and $c^+$,
which are obtained when we attempt to apply the argument used in the
proof of the last internal cut lemma in order to construct an overall
latest cut.}
\label {Fig.22}
\end {figure}

\begin {figure}
\vspace{50 mm}
\hskip 22 mm
\special {illustration :Diagrams:Fig.23}
\vskip 3 mm
\caption {The two possible ``latest" cuts, $c_1$ and $c_2$, which lead
to the double-counting of Fig.~
\protect{\ref{Fig.21}}, and the ``cuts" which are obtained when we
attempt to apply the argument used in the proof of the last internal
cut lemma in order to construct an overall latest cut.}
\label {Fig.23}
\end {figure}

\vfill
\eject
\null

\begin {figure}
\vspace{70 mm}
\hskip 15 mm
\special {illustration :Diagrams:Fig.24}
\vskip 3 mm
\caption {The sum of $\ctt{j} \cap \ctt{k}$ in diagrammatic form, given
that $2t_i \geq m$ and $(m/\tit{j}) \cap (m/\tit{k}) = \phi$.}
\label {Fig.24}
\end {figure}

\begin {figure}
\vspace{100 mm}
\hskip -5 mm
\special {illustration :Diagrams:Fig.25}
\vskip 3 mm
\caption {If the cut $Y$ is an $r$-cut then it is clear that  the cut
$Y'$ will cut less than $r$ lines.}
\label {Fig.25}
\end {figure}

\vfill
\eject
\null

\begin {figure}
\vspace{85 mm}
\hskip 25 mm
\special {illustration :Diagrams:Fig.26 scaled 800}
\vskip 3 mm
\caption {The sum of the intersection $\cfost{j} \cap \cfost{k}$ in the case
$2t_i \geq m$ and $2s_f < n$, given that $(m/\tit{j}) \cap (m/\tit{k}) = \phi$
and $\sft{j} \cap \sft{k}=\phi$.}
\label {Fig.26}
\end {figure}

\begin {figure}
\vspace{85 mm}
\hskip 25 mm
\special {illustration :Diagrams:Fig.27 scaled 800}
\vskip 3 mm
\caption {The sum of the intersection $\cfost{j} \cap \cfost{k}$ in the case
$2t_i < m$ and $2s_f \geq n$, given that $(n/\sft{j}) \cap (n/\sft{k}) = \phi$
and $\tit{j} \cap \tit{k}=\phi$.}
\label {Fig.27}
\end {figure}

\vfill
\eject
\null

\begin {figure}
\vspace{55 mm}
\hskip 20 mm
\special {illustration :Diagrams:Fig.28 scaled 800}
\vskip 3 mm
\caption {The sum of sub-sub-class $\cfost{j}$, and the $r$-cut $X$ which may
be made to obtain this sum, with the $r$-cut $Y$ which indicates that all
diagrams in this sub-sub-class are double-counted in the case $2s_f=n$ and
$2t_i=m$.}
\label {Fig.28}
\end {figure}

\begin {figure}
\vspace{85 mm}
\hskip 35 mm
\special {illustration :Diagrams:Fig.29 scaled 700}
\vskip 3 mm
\caption {The diagram which both $\cfost{j}$ and $\cfost{k}$ sum to, given that
$2s_f=n$, $2t_i=m$, $\sft{j} \cap \sft{k}=\phi$ and $\tit{j}
\cap \tit{k}=\phi$.}
\label {Fig.29}
\end {figure}

\vfill
\eject
\null

\begin {figure}
\vspace{85 mm}
\hskip 35 mm
\special {illustration :Diagrams:Fig.30 scaled 800}
\vskip 3 mm
\caption {The sum of the intersection $\cfs{j} \cap \cfs{k}$ in diagrammatic
form if $2s_f \geq n$ and $(n/\sft{j}) \cap (n/\sft{k}) = \phi$.}
\label {Fig.30}
\end {figure}

\end {document}